\documentclass{aastex61}
\received{7 December 2018}
\revised{20 February 2019}
\accepted{14 April 2019}
\submitjournal{AJ}

\shorttitle{Qatar-8b, 9b and 10b exoplanets}
\shortauthors{Alsubai et al.}

\usepackage{amsmath,texlogos}
\usepackage{natbib}
\usepackage{subfigure}
\usepackage{graphicx}
\usepackage{txfonts}
\usepackage[T1]{fontenc}
\usepackage{aecompl}
\usepackage{times}

\newcommand{\kms}{\ensuremath{\rm km\,s^{-1}}}
\newcommand{\ms}{\ensuremath{\rm m\,s^{-1}}}
\newcommand{\logg}{\ensuremath{\log{g}}}

\newcommand{\bjdtdb}{\ensuremath{\rm {BJD_{TDB}}}}
\newcommand{\feh}{\ensuremath{\left[{\rm Fe}/{\rm H}\right]}}
\newcommand{\mh}{\ensuremath{\left[m/{\rm H}\right]}}
\newcommand{\teff}{\ensuremath{T_{\rm eff}}}
\newcommand{\teq}{\ensuremath{T_{\rm eq}}}

\newcommand{\msun}{\ensuremath{{\,M_\odot}}}
\newcommand{\rsun}{\ensuremath{{\,R_\odot}}}
\newcommand{\lsun}{\ensuremath{{\,L_\odot}}}

\newcommand{\mstar}{\ensuremath{\,M_\star}}
\newcommand{\rstar}{\ensuremath{\,R_\star}}

\newcommand{\mpl}{\ensuremath{\,M_{\rm P}}}
\newcommand{\rpl}{\ensuremath{\,R_{\rm P}}}
\newcommand{\mj}{\ensuremath{{\,M_{\rm J}}}}
\newcommand{\rj}{\ensuremath{{\,R_{\rm J}}}}

\begin{document}

\title{Qatar Exoplanet Survey: Qatar-8b, 9b and 10b --- A Hot Saturn and Two Hot Jupiters}

\correspondingauthor{Khalid Alsubai}
\email{kalsubai@qf.org.qa}

\author{Khalid Alsubai}
\affil{Hamad bin Khalifa University (HBKU), Qatar Foundation, PO Box 5825, Doha, Qatar}

\author{Zlatan I. Tsvetanov}
\affiliation{Hamad bin Khalifa University (HBKU), Qatar Foundation, PO Box 5825, Doha, Qatar}

\author{Stylianos Pyrzas}
\affiliation{Hamad bin Khalifa University (HBKU), Qatar Foundation, PO Box 5825, Doha, Qatar}

\author{David W. Latham}
\affiliation{Harvard-Smitsonian Center for Astrophysics, 60 Garden Street,  Cambridge, MA 02138, USA}

\author{Allyson Bieryla}
\affiliation{Harvard-Smitsonian Center for Astrophysics, 60 Garden Street,  Cambridge, MA 02138, USA}

\author{Jason Eastman}
\affiliation{Harvard-Smitsonian Center for Astrophysics, 60 Garden Street,  Cambridge, MA 02138, USA}

\author{Dimitris Mislis}
\affiliation{Hamad bin Khalifa University (HBKU), Qatar Foundation, PO Box 5825, Doha, Qatar}

\author{Gilbert A. Esquerdo}
\affiliation{Harvard-Smitsonian Center for Astrophysics, 60 Garden Street,  Cambridge, MA 02138, USA}

\author{John Southworth}
\affiliation{Astrophysics Group, Keele University, Staffordshire ST5 5BG, UK}

\author{Luigi Mancini}
\affiliation{Department of Physics, University of Rome Tor Vergata, Via della Ricerca Scientifica 1, I-00133 Roma, Italy}
\affiliation{Max Planck Institute for Astronomy, K\"{o}nigstuhl 17, D-69117 Heidelberg, Germany}
\affiliation{INAF-Osservatorio Astrofisico di Torino, Via Osservatorio 20, I-10025 Pino Torinese, Italy}
\affiliation{International Institute for Advanced Scientific Studies (IIASS), Via G. Pellegrino 19, I-84019 Vietri sul Mare (SA), Italy}

\author{Ali Esamdin}
\affiliation{Xinjiang Astronomical Observatory (XAO), Chinese Academy of Sciences, 150 Science 1-Street, Urumqi, Xinjiang 830011, China }

\author{Jinzhong Liu}
\affiliation{Xinjiang Astronomical Observatory (XAO), Chinese Academy of Sciences, 150 Science 1-Street, Urumqi, Xinjiang 830011, China }

\author{Lu Ma}
\affiliation{Xinjiang Astronomical Observatory (XAO), Chinese Academy of Sciences, 150 Science 1-Street, Urumqi, Xinjiang 830011, China }

\author{Marc Bretton}
\affiliation{Observatoire des Baronnies Proven\c{c}ales (OBP), Le Mas des Gr\'{e}s, Route de Nyons, 05150 Moydans, France }

\author{Enric Pall\'{e}}
\affiliation{Instituo de Astrof\'{i}sica da Canarias (IAC), 38205 La Laguna, Tenerife, Spain}
\affiliation{Departamento de Astrof\'{i}sica, Universidad de La Laguna (ULL), 38206 La Laguna, Tenerife, Spain}

\author{Felipe Murgas}
\affiliation{Instituo de Astrof\'{i}sica da Canarias (IAC), 38205 La Laguna, Tenerife, Spain}
\affiliation{Departamento de Astrof\'{i}sica, Universidad de La Laguna (ULL), 38206 La Laguna, Tenerife, Spain}

\author{Nicolas P. E. Vilchez}
\affiliation{Hamad bin Khalifa University (HBKU), Qatar Foundation, PO Box 5825, Doha, Qatar}

\author{Hannu Parviainen}
\affiliation{Instituo de Astrof\'{i}sica da Canarias (IAC), 38205 La Laguna, Tenerife, Spain}
\affiliation{Departamento de Astrof\'{i}sica, Universidad de La Laguna (ULL), 38206 La Laguna, Tenerife, Spain}

\author{Pilar Monta\~{n}es-Rodriguez}
\affiliation{Instituo de Astrof\'{i}sica da Canarias (IAC), 38205 La Laguna, Tenerife, Spain}
\affiliation{Departamento de Astrof\'{i}sica, Universidad de La Laguna (ULL), 38206 La Laguna, Tenerife, Spain}

\author{Norio Narita}
\affiliation{Department of Astronomy, Graduate School of Science, The University of Tokyo, 7-3-1 Hongo, Bunkyo-ku, Tokyo 113-0033, Japan}
\affiliation{Astrobiology Center, National Institutes of Natural Sciences, 2-21-1 Osawa, Mitaka, Tokyo 181-8588, Japan}
\affiliation{JST, PRESTO, 7-3-1 Hongo, Bunkyo-ku, Tokyo 113-0033, Japan}
\affiliation{National Astronomical Observatory of Japan, 2-21-1 Osawa, Mitaka, Tokyo 181-8588, Japan}
\affiliation{Instituo de Astrof\'{i}sica da Canarias (IAC), 38205 La Laguna, Tenerife, Spain}

\author{Akihiko Fukui}
\affiliation{Department of Earth and Planetary Science, Graduate School of Science, The University of Tokyo, 7-3-1 Hongo, Bunkyo-ku, Tokyo 113-0033, Japan}
\affiliation{Instituo de Astrof\'{i}sica da Canarias (IAC), 38205 La Laguna, Tenerife, Spain}

\author{Nobuhiko Kusakabe}
\affiliation{Astrobiology Center, National Institutes of Natural Sciences, 2-21-1 Osawa, Mitaka, Tokyo 181-8588, Japan}

\author{Motohide Tamura}
\affiliation{Department of Astronomy, Graduate School of Science, The University of Tokyo, 7-3-1 Hongo, Bunkyo-ku, Tokyo 113-0033, Japan}
\affiliation{Astrobiology Center, National Institutes of Natural Sciences, 2-21-1 Osawa, Mitaka, Tokyo 181-8588, Japan}

\author{Khalid Barkaoui}
\affiliation{Space sciences, Technologies and Astrophysics Research (STAR) Institute, Universit\'e de Li\`ege, All\'ee du 6 Ao\^{u}t 17, Bat. B5C, 4000, Li\`ege, Belgium}
\affiliation{Oukaimeden Observatory, High Energy Physics and Astrophysics Laboratory, Cadi Ayyad University, Marrakech, Morocco}

\author{Francisco Pozuelos}
\affiliation{Space sciences, Technologies and Astrophysics Research (STAR) Institute, Universit\'e de Li\`ege, All\'ee du 6 Ao\^{u}t 17, Bat. B5C, 4000, Li\`ege, Belgium}

\author{Michael Gillon}
\affiliation{Space sciences, Technologies and Astrophysics Research (STAR) Institute, Universit\'e de Li\`ege, All\'ee du 6 Ao\^{u}t 17, Bat. B5C, 4000, Li\`ege, Belgium}

\author{Emmanuel Jehin}
\affiliation{Space sciences, Technologies and Astrophysics Research (STAR) Institute, Universit\'e de Li\`ege, All\'ee du 6 Ao\^{u}t 17, Bat. B5C, 4000, Li\`ege, Belgium}

\author{Zouhair Benkhaldoun}
\affiliation{Oukaimeden Observatory, High Energy Physics and Astrophysics Laboratory, Cadi Ayyad University, Marrakech, Morocco}

\author{Ahmed Daassou}
\affiliation{Oukaimeden Observatory, High Energy Physics and Astrophysics Laboratory, Cadi Ayyad University, Marrakech, Morocco}

\author{Hani M. Dalee}
\affiliation{Hamad bin Khalifa University (HBKU), Qatar Foundation, PO Box 5825, Doha, Qatar}




\begin{abstract}
In this paper we present three new extrasolar planets from the Qatar Exoplanet 
Survey (QES). Qatar-8b is a hot Saturn, with $\mpl = 0.37\mj$ and $\rpl =1.3\rj$, 
orbiting a solar-like star every $P_{orb}=3.7$ days. Qatar-9b is a hot Jupiter with 
a mass of $\mpl = 1.2\mj$ and a radius of $\rpl =1\rj$, in a $P_{orb}=1.5$-days 
orbit around a low mass, $\mstar = 0.7\msun$, mid-K main-sequence star. 
Finally, Qatar-10b is a hot, $\teq \sim2000$ K, sub-Jupiter mass planet, 
$\mpl =0.7\mj$, with a radius of $\rpl = 1.54\rj$ and an orbital period of 
$P_{orb}=1.6$ days, placing it on the edge of the sub-Jupiter desert.

\end{abstract}




\section{Introduction} \label{sec:Introduction}

Since \citet{51peg} announced 51\,Peg, the first extrasolar planet around a 
main-sequence star, the number of extrasolar planets has been rising steadily, 
revealing the large diversity in physical properties and configurations of the 
underlying extrasolar planets population. In order to properly understand this 
diversity, a large  sample of well-characterized, in terms of physical properties, 
planets and their respective host stars is required.

Large-scale, ground-based surveys for transiting extrasolar planets, such as 
OGLE-III \citep{OGLE}; TrES \citep{TrES}; HATNet \citep{HAT}; XO \citep{XO}; 
WASP \citep{WASP}; KELT \citep{KELT}; and QES \citep{alsubai2013}, have 
played a pivotal role both in significantly increasing the numbers of known planets 
and in providing prime targets to fulfill the ``well-characterized'' requirement. By 
design, these surveys offer certain advantages: (i) the very fact that the planets 
are transiting implies that, generally, both the actual mass ($M_P$, not only 
$M_P\mathrm{sini}\,i$) and the planet radius (and by extension, the bulk density) 
can be determined; (ii) ground-based surveys are more sensitive to brighter host 
stars and larger planets; this (usually) allows for the physical properties of the 
planet to be determined with good precision (better than 10\%) and offers the 
possibility of individual systems suitable for intensive follow-up studies.

In this paper we present three new transiting extrasolar planets discovered 
by QES: Qatar-8b --- a hot Saturn around a solar-like star; Qatar-9b -- a hot 
Jupiter orbiting a mid-K main-sequence star; and Qatar-10b -- a hot Jupiter 
around a late-F main-sequence star. The paper is organized as follows: in 
Section \ref{sec:Observations} we present the survey photometry and 
describe the follow-up photometry and spectroscopy used to confirm the 
planetary nature of the transits. In Section \ref{sec:Analysis} we present the 
analysis of the data and the global system solutions using simultaneous 
fits to the available radial velocities (RVs) and follow-up photometric light 
curves, and in Section \ref{sec:conclusions} we summarize our results and 
put the three new planets in the broader context of the exoplanets field. 

\section{Observations} \label{sec:Observations} 

\subsection{Discovery photometry} \label{subsec:DiscPhot}

The survey data were collected with QES, hosted by the New Mexico Skies 
Observatory\footnote{http://www.nmskies.com} located in Mayhill, NM, USA. A full 
description of QES can be found in our previous publications, e.g., \cite{alsubai2013}, 
\cite{alsubai2017}. 

The discovery light curves of Qatar-8b and Qatar-9b contain 2\,959 and 2\,755 data 
points, respectively, obtained during observations from December 5, 2016 to May 9, 
2017. For Qatar-10b, the discovery light curve has 2\,077 data points collected in the 
time period March 21 -- November 1, 2017. The survey images are run through the QES 
pipeline, which extracts the photometric measurements using the image subtraction 
algorithm by \cite{dbdia}. A full description of the pipeline can be found in \cite{alsubai2013}. 

The output light curves are ingested into the QES archive and are detrended using a 
combination of the \texttt{Trend Filtering Algorithm} (TFA, \citealt{kovacs1}), which 
constructs a filter function from a set of field stars considered to be a representative 
template for systematics in the field, and the \texttt{DOHA} algorithm \citep{mislis}, 
a co-trending algorithm used to eliminate lingering, quasi-systematic patterns 
identified from groups of stars that are highly correlated to each other. The light 
curves are then further processed with the \texttt{TSARDI} algorithm \citep{tsardi}, 
a machine learning points rejection algorithm that deals with any residual data 
irregularities. Qatar-8b, 9b and 10b were identified as strong candidates during a 
search for transit-like events using the \texttt{Box Least Squares} algorithm (BLS) of 
\cite{kovacs2}, following a procedure similar to that described in \cite{collier}. Note 
that, although the initial candidate selection is an automatic procedure, the final vetting 
is done by eye. 

\begin{figure*}
\centering
\includegraphics[width=8.8cm]{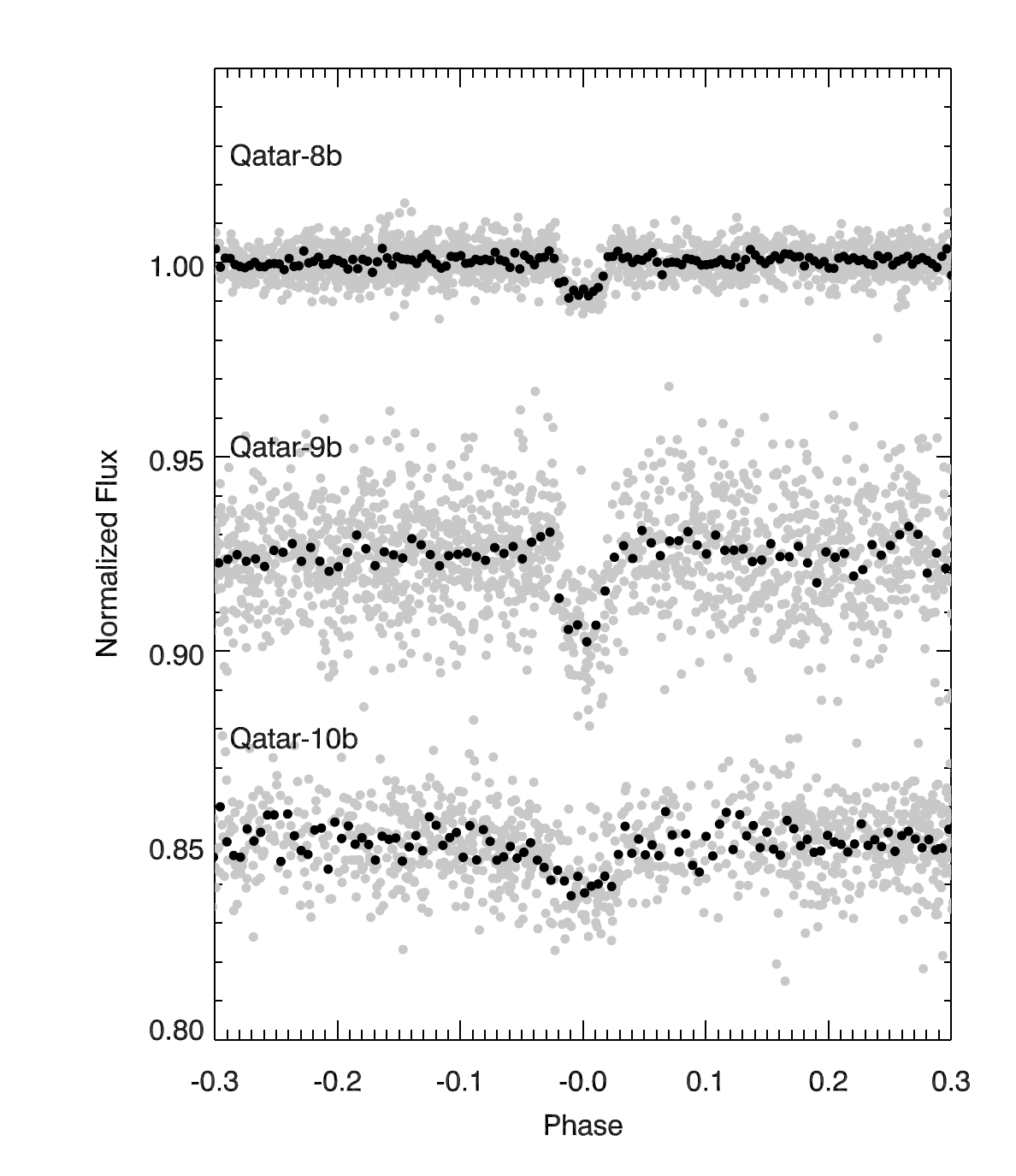}
\caption{The discovery light curves for Qatar-8b ({\it top}), Qatar-9b ({\it middle}) and 
Qatar-10b ({\it bottom}) folded to the period identified by the BLS analysis and plotted 
with an arbitrary vertical offset for clarity. The gray points represent the original 
observations, and the black points are the binned values to better guide the eye.}
\label{fig:q8910_discovery}
\end{figure*}

Figure\,\ref{fig:q8910_discovery} shows the discovery light curves for the three 
exoplanets discussed in this paper.

\subsection{Follow-up photometry} \label{subsec:FollowPhot}

Follow-up photometric observations of a number of transits of Qatar-8b, 9b and 10b 
were collected at five different observatories with the following combination of 
telescopes and instruments: 

{FLWO:} The 1.5\,m telescope at the Fred L.\,Whipple Observatory (Mount Hopkins, 
Arizona, USA) in combination with KeplerCam, equipped with a single 4k$\times$4k 
Fairchild CCD with a 0\farcs37 pixel$^{-1}$ and a $23\farcm1 \times 23\farcm1$ 
on-sky field of view (FOV).

{QFT:} The 0.5\,m Qatar Follow-up Telescope (New Mexico Skies Observatory, Mayhill, 
New Mexico, USA), equipped with a 1k$\times$1k Andor iKon-M 934 CCD, yielding a 
FOV of $13\arcmin \times 13\arcmin$. 
 
{OBP:} The 0.82\,m telescope at the Observatoire des Baronnies 
Proven\c{c}ales\footnote{http://www.obs-bp.fr} (Provence-Alpes-C\^{o}te d'Azur, France) 
equipped with a FLI ProLine PL230 camera with a 2k$\times$2k e2v CCD detector 
resulting in a $23\arcmin \times 23\arcmin$ FOV.
 
{CAHA}: The 1.23\,m Zeiss telescope at the Centro Astron\'{o}mico Hispano-Alem\'{a}n 
(Calar Alto, Spain) in combination with the DLR-MKIII camera with a 4k$\times$4k, e2v 
CCD resulting in a $21\farcm5 \times 21\farcm5$ FOV.
 
{TCS:} The 1.52\,m Telescopio Carlos Sanchez at the Teide Observatory (Tenerife, 
Canary Islands, Spain) with the MuSCAT2 instrument, which takes images in 4 filters 
simultaneously. Each channel is equipped with a 1k$\times$1k CCD, resulting in a  
$7\farcm4 \times 7\farcm4$ on-sky FOV. For a detailed description of MuSCAT2 and its 
dedicated photometric pipeline see \cite{MuSCAT2}.
 
{TRAPPIST-North}: The 0.6\,m robotic TRansiting Planets and PlanetesImals Small 
Telescope\footnote{https://www.trappist.uliege.be} is located at Oukaimeden Observatory, 
Morocco. It is equipped with a 2k$\times$2k deep-depletion Andor IKONL BEX2 DD 
CCD camera with a pixel scale of 0\farcs60 and an on-sky FOV of 
$19\farcm8 \times 19\farcm8$.

All follow-up light curves were generated through differential aperture photometry 
performed on the sequence of images for each observing run. In each case a 
number of comparison stars were selected and those with excessive noise or 
suspected variability were excluded from the final analysis. For the CAHA 
observations, the telescope was defocused and data reduction was carried out 
using the \texttt{DEFOT} pipeline \citep{Southworth2009, Southworth2014}. For 
observations taken at FLWO, OBP and with QFT the telescope was kept only 
approximately in focus and we used the AstroImageJ software package \citep{AIJ} 
to extract the light curves. The MuSCAT2 instrument on TCS has a dedicated pipeline 
for extracting the light curves. The extraction of the fluxes for the TRAPPIST-North 
observations was done by aperture photometry on selected stars with the 
IRAF/DAOPHOT software package \citep{daophot}. Final transit light curves 
were produced by normalizing to a low order polynomial (maximum order 2 
in one case, a straight line for all other cases) fitted to the flat part of the light 
curves which also removes small residual trends if present. The uncertainties 
in the CAHA and MuSCAT2 light curves are estimated from the point-to-point 
dispersion of the points out of transit and for all other cases are the combination 
of the photon noise and the background noise (see the AIJ package, \citealt{AIJ}). 
Nevertheless, as the main source of uncertainties is residual systematics and 
not photon noise, when fitting the follow-up light curves EXOFASTv2 (see Section
\ref{subsec:GlobalFit}) adds a variance term to each transit to enforce reduced 
$\chi^{2} \sim 1$. 

\begin{figure*}
\centering
\includegraphics[width=8.8cm]{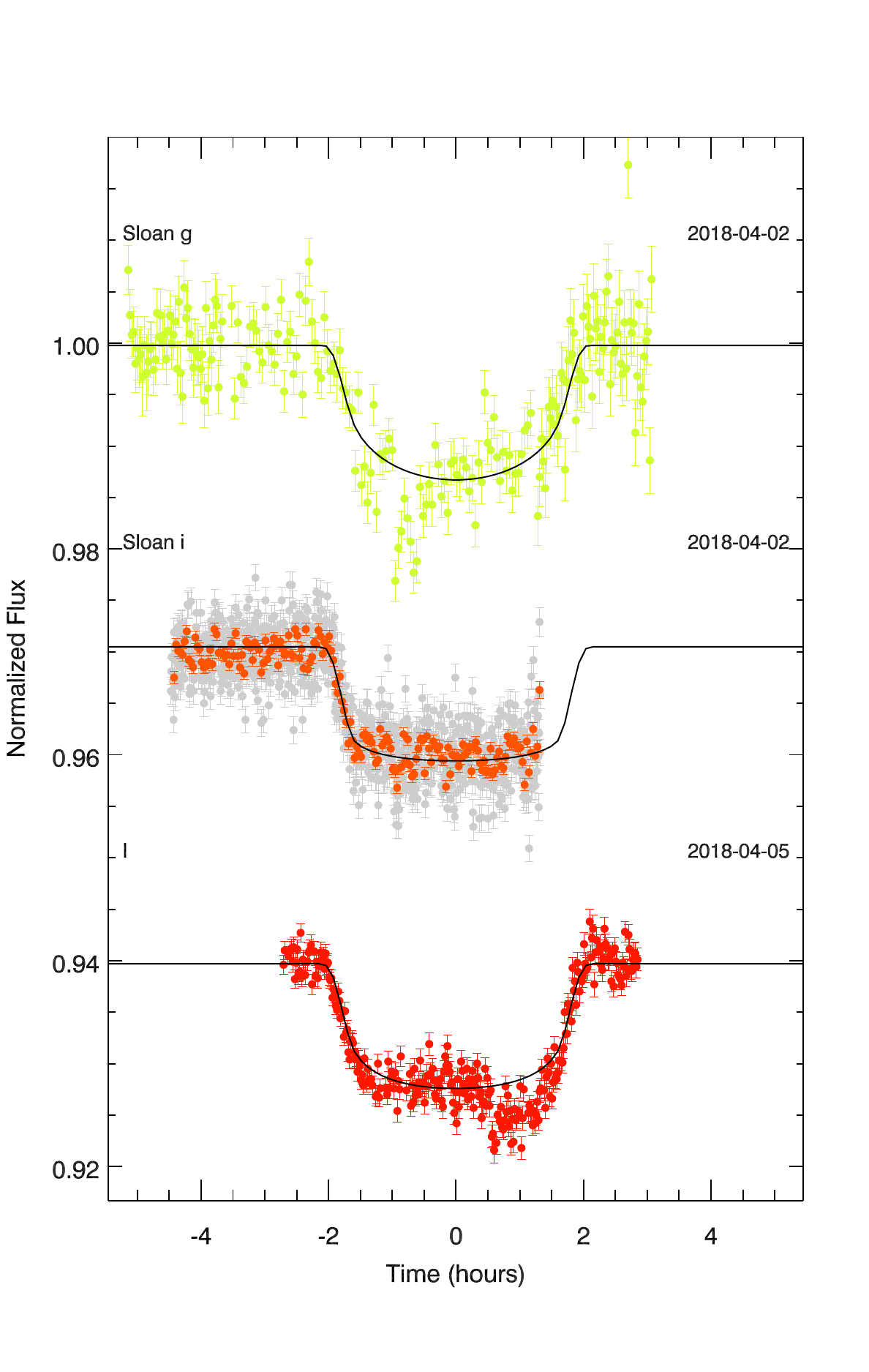}
\includegraphics[width=8.8cm]{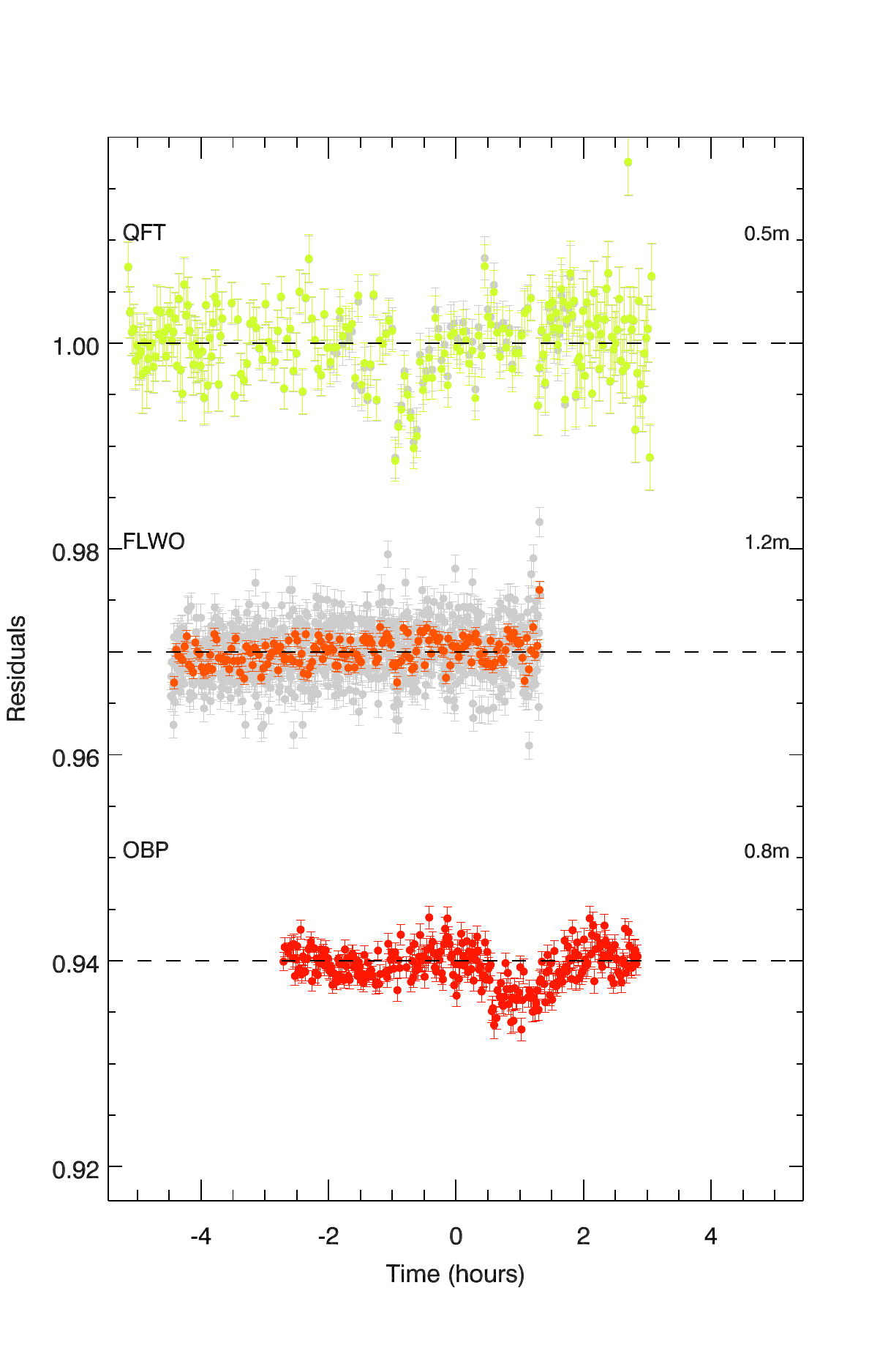}
\caption{The three follow-up light curves of Qatar-8b. The {\it left} panel shows the 
light curves ordered from top to bottom as they appear in Table\,\ref{table:FollowUpPhotLog} 
with a vertically added shift for clarity. The solid, black lines are the best model fits (see 
Section\,\ref{subsec:GlobalFit}). The residuals from the fits are shown in the 
{\it right} panel. The individual data points are color coded according to the 
filter used and for observations taken with KeplerCam and Muscat2 we show both 
the original data points (light gray) as well as the data binned to a uniform cadence 
of 2 min. The filter, date of observation, observatory and telescope size are also 
given in the two panels.}
\label{fig:q8bTR}
\end{figure*}

\begin{figure*}
\centering
\includegraphics[width=8.8cm]{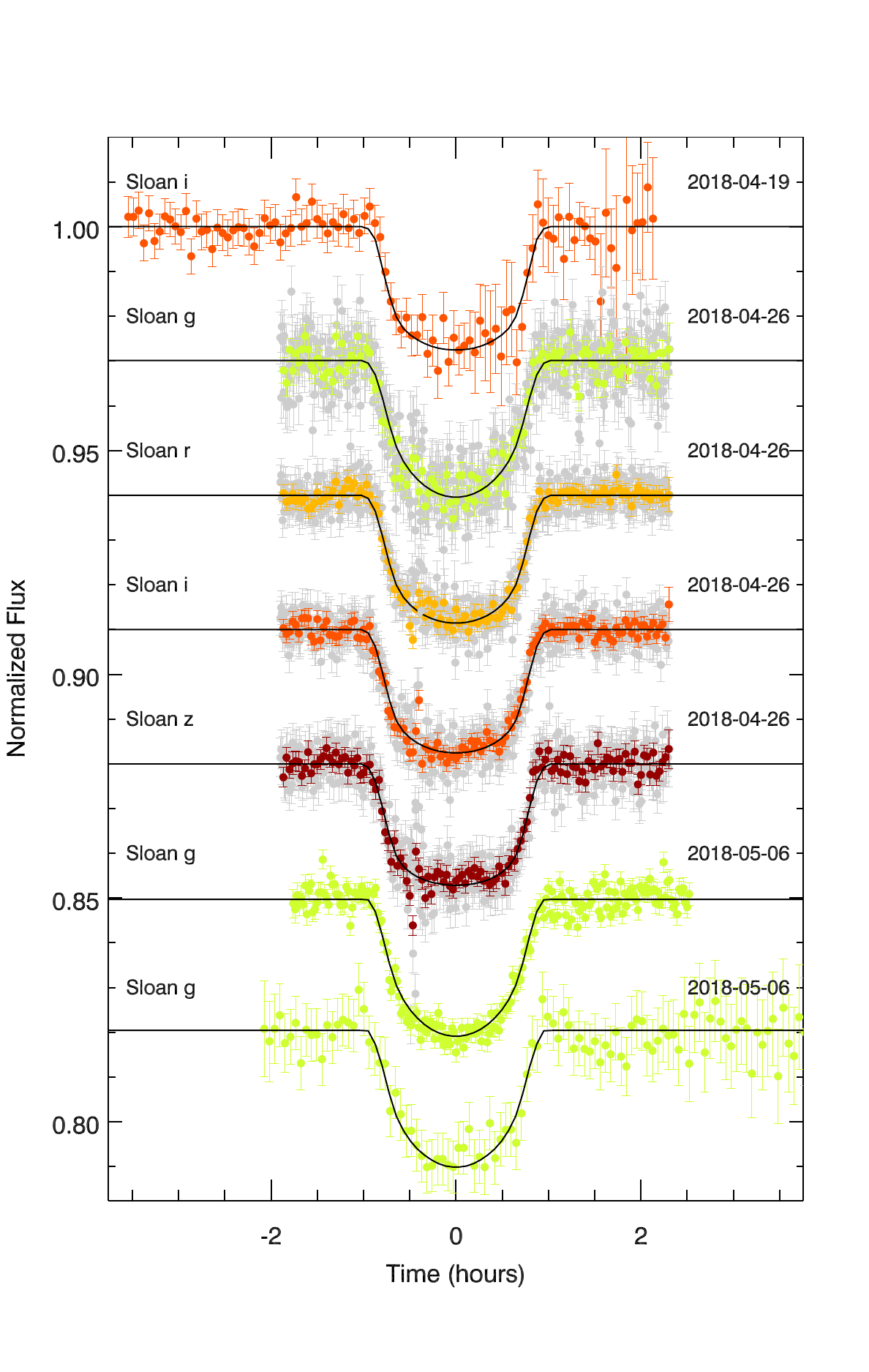}
\includegraphics[width=8.8cm]{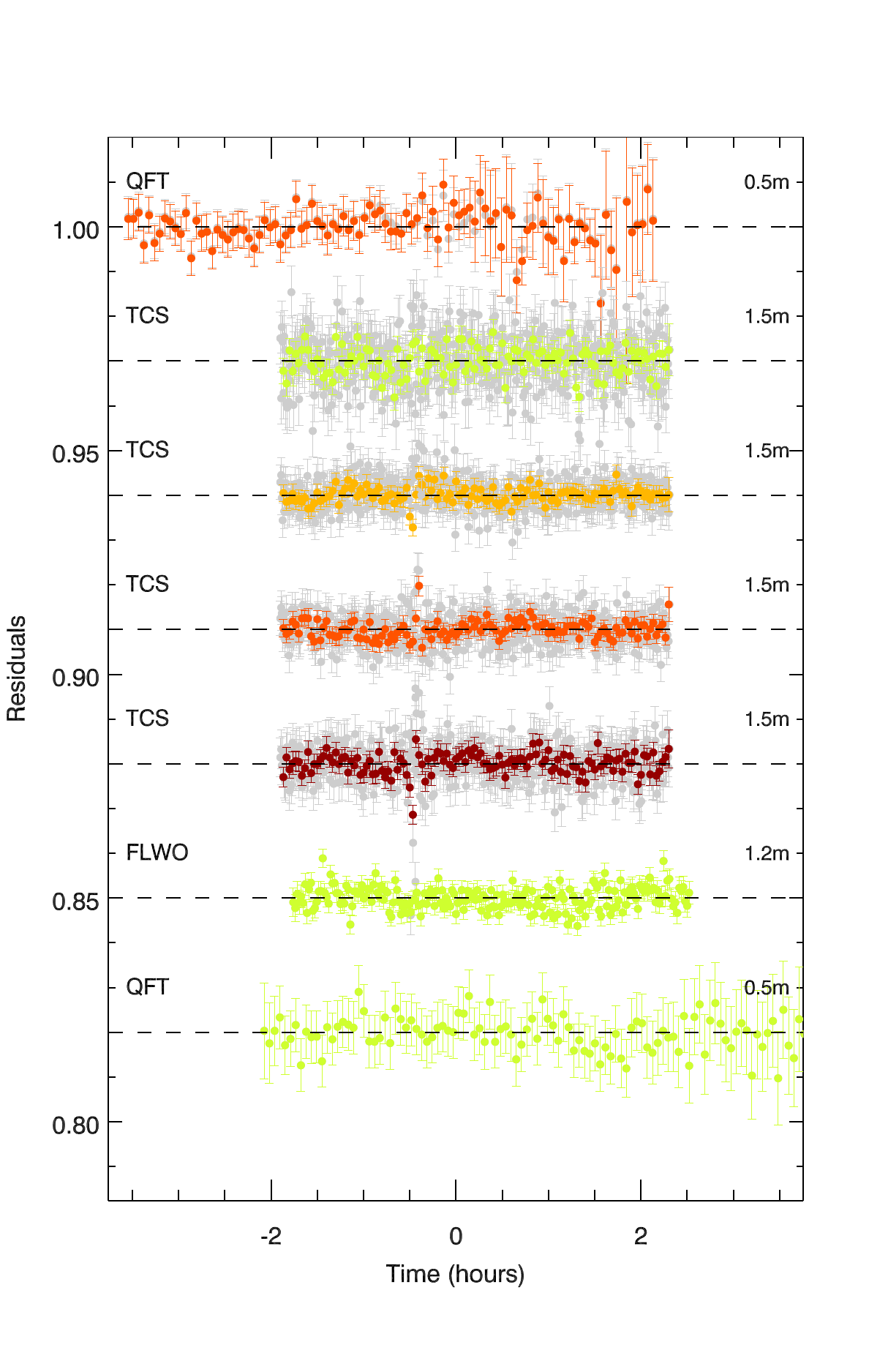}
\caption{The seven follow-up transit light curves of Qatar-9b. All symbols, labels and colors 
follow the same convention as in Figure \ref{fig:q8bTR}.}
\label{fig:q9bTR}
\end{figure*}

\begin{figure*}
\centering
\includegraphics[width=8.8cm]{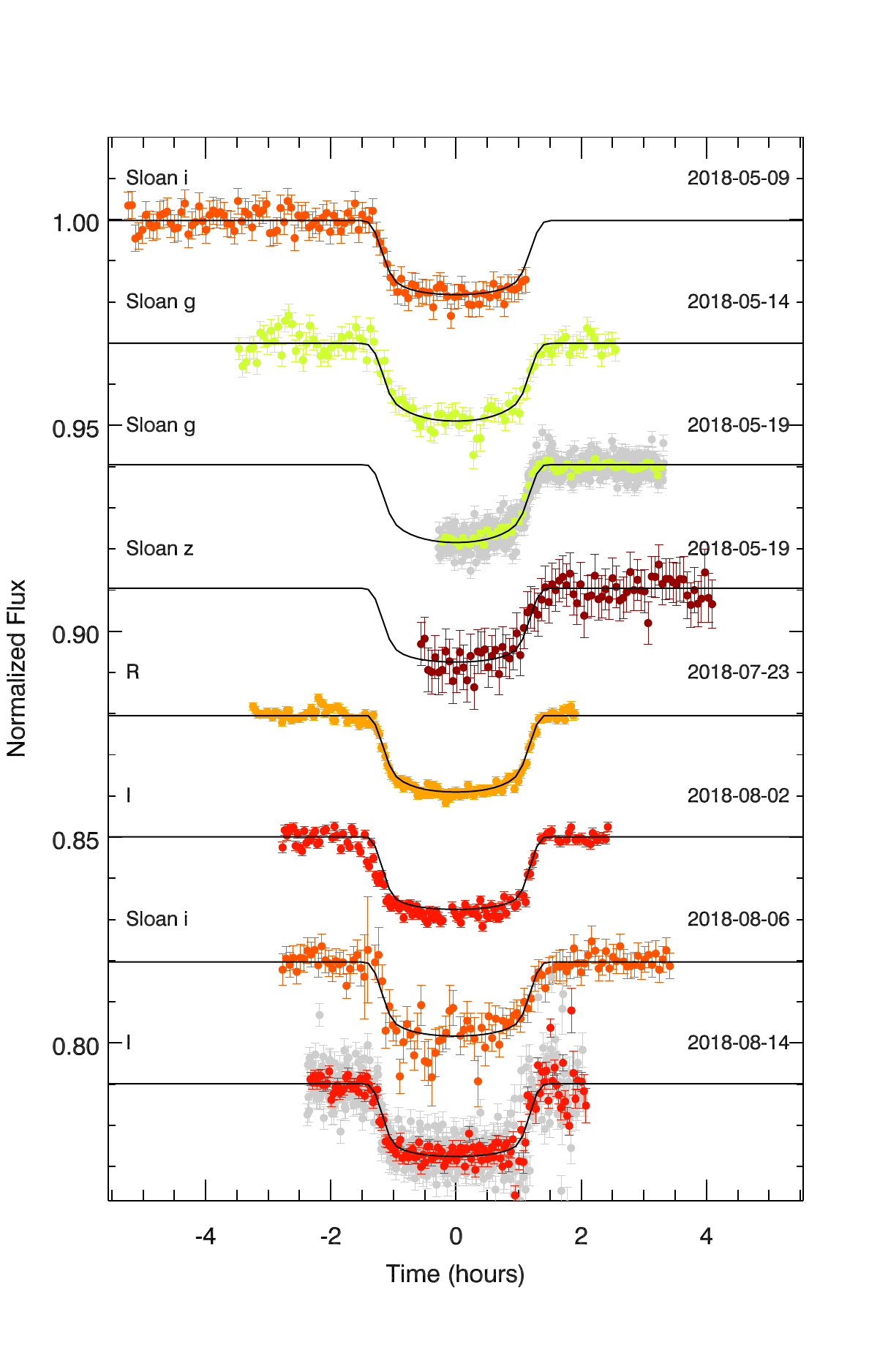}
\includegraphics[width=8.8cm]{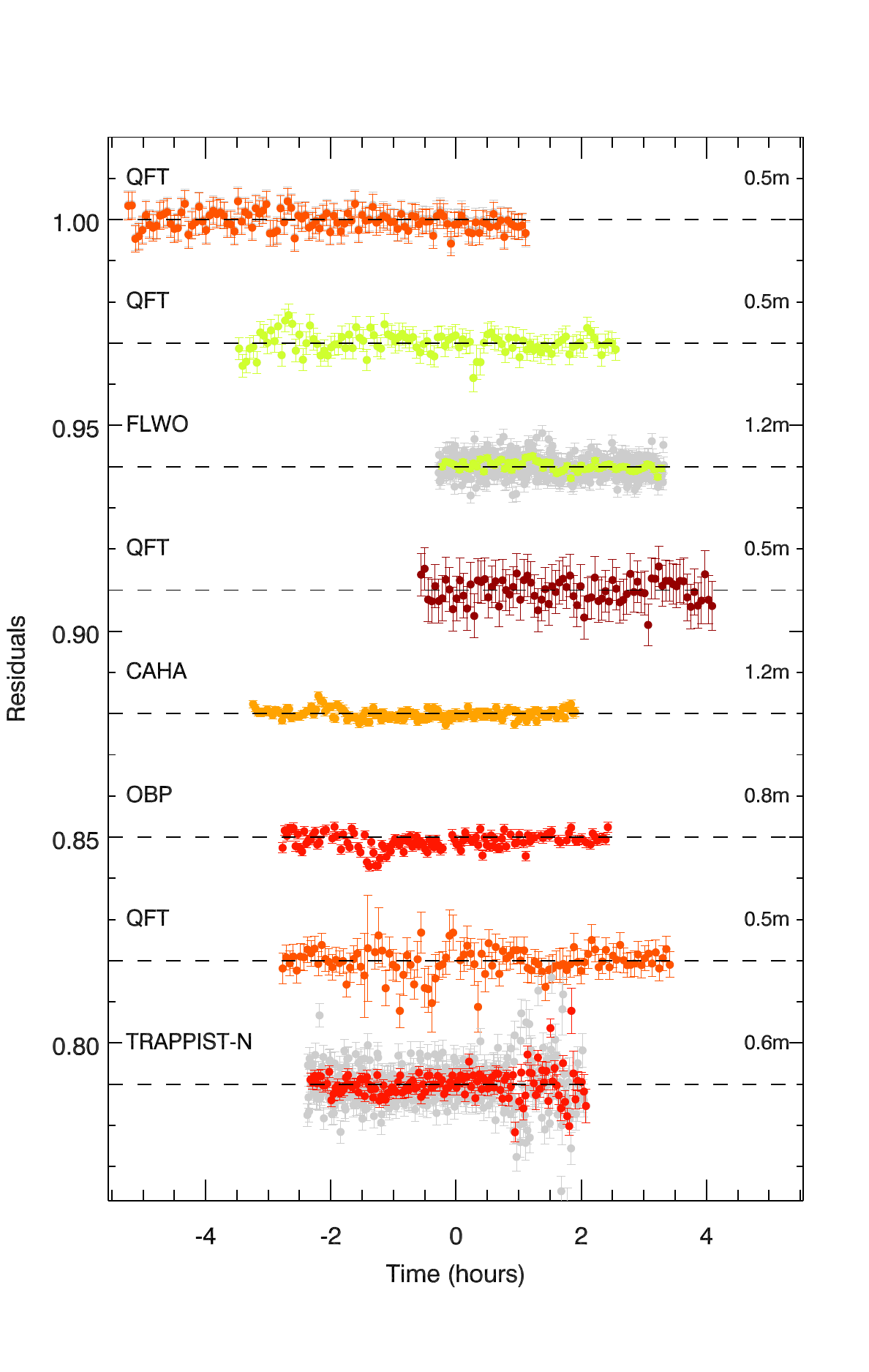}
\caption{Same as Figures \ref{fig:q8bTR} and \ref{fig:q9bTR} but for Qatar-10b. All symbols, 
labels and colors follow the same convention.}
\label{fig:q10bTR}
\end{figure*}

A summary of our follow-up photometric observations is given in Table 
\ref{table:FollowUpPhotLog} where we list the date, telescope, filter, and
cadence for each transit observation. In the last column we give the mean 
uncertainty in 2 min bins if the cadence is less than 2 min, otherwise for 
the original observations cadence. The resulting light curves, along with the 
best model fit and the corresponding residuals, are plotted in 
Figures \ref{fig:q8bTR}, \ref{fig:q9bTR} and \ref{fig:q10bTR}.

\begin{table}
\centering
\caption{Log of follow-up photometric observations for Qatar-8b, 9b and 10b. See 
text for details on telescopes and instruments. The last columns gives the mean 
error in 2 min bins.}
\label{table:FollowUpPhotLog}
\begin{tabular}{lllcll}
\hline\hline
ID & Date & Telescope & Filter & Cadence, s & RMS, mmag \\
\hline
\multicolumn{2}{l}{Qatar-8b} & & & & \\
1 & 2018-04-02 &  QFT & g & 35 & 2.3 \\
2 & 2018-04-02 &  FLWO & i & 30 & 0.6 \\
3 & 2018-04-05 &  OBP & I & 125 & 1.6 \\
\multicolumn{2}{l}{Qatar-9b} & & & & \\
1 & 2018-04-19 & QFT & i & 204 & 4.2 \\
2 & 2018-04-26 & TCS & g & 30 & 2.9 \\
3 & 2018-04-26 & TCS & r & 30 & 1.9 \\
4 & 2018-04-26 & TCS & i & 30 & 1.9 \\
5 & 2018-04-26 & TCS & z & 30 & 2.2 \\
6 & 2018-05-06 & FLWO & g & 35 & 1.5 \\
7 & 2018-05-06 & QFT & g & 204 & 6.0 \\
\multicolumn{2}{l}{Qatar-10b} & & & & \\
1 & 2018-05-09 & QFT & i & 204 & 3.0 \\
2 & 2018-05-14 & QFT & g & 204 & 2.6 \\
3 & 2018-05-19 & QFT & z & 204 & 5.1 \\
4 & 2018-05-19 & FLWO & g & 28 & 0.9 \\
5 & 2018-07-23 & CAHA & R & 116 & 1.0 \\
6 & 2018-08-02 & OBP & I & 124 & 1.1 \\
7 & 2018-08-06 & QFT & i & 204 & 3.2 \\
8 & 2018-08-14 & TRAPPIST-N & I & 32 & 1.7 \\
\hline
\end{tabular}
\end{table}

\subsection{Follow-up spectroscopy} \label{subsec:FollowSpec}

Follow-up spectroscopic observations to measure precision radial velocities for all 
three targets -- Qatar-8b, 9b, and 10b -- were obtained in the same manner as for 
all previous QES candidates (for details see \citet{alsubai2011} and subsequent 
papers). In brief, we used the Tillinghast Reflector Echelle Spectrograph (TRES) on 
the 1.5\,m Tillinghast Reflector at FLWO. All spectra were obtained using the medium 
fiber, which results in a resolving power of $R \sim$ 44,000 and a velocity resolution 
element of 6.8 \kms\ FWHM. The wavelength calibration was established using a 
Th-Ar hollow-cathode lamp illuminating the science fiber with two exposures obtained 
immediately before and after each target spectrum. 

\begin{table}
\centering
\caption{Relative RVs and BS variations for Qatar-8.}
\label{table:q8bRV}
\begin{tabular}{lrr}
\hline\hline
BJD$_{TDB}$       &RV (\ms)          &BS (\ms)       \\
\hline
TRES & & \\
2458183.747354 & $   22.6 \pm 31.4 $ & $ 12.7 \pm 14.5 $ \\
2458207.782751 & $-113.5 \pm 27.0 $ & $-28.7 \pm   8.8 $ \\
2458211.702819 & $ -59.5 \pm 30.0 $ &  $   3.0 \pm 15.5 $ \\
2458218.684754 & $ -65.5 \pm 25.2 $ &  $   6.3 \pm 12.3 $ \\
2458226.645789 & $ -77.7 \pm 37.6 $ &  $ 35.1 \pm 14.8 $ \\
2458228.646936 & $  20.7 \pm 26.3 $ &  $ 35.8 \pm 14.5 $ \\
2458241.668589 & $ -80.9 \pm 37.2 $ &  $-12.3 \pm 15.8 $ \\
2458244.754532 & $ -57.5 \pm 32.6 $ &  $-32.1 \pm 21.5 $ \\
2458261.706560 & $ 	  0.0 \pm 27.0 $ &  $ 12.2 \pm   8.0 $ \\
2458263.738688 & $-110.5 \pm 24.5 $ & $-24.9 \pm 17.4 $ \\
2458267.667690 & $ -81.7 \pm 20.4 $ &  $-34.7 \pm 15.7 $ \\
2458274.674490 & $ -64.9 \pm 33.0 $ &  $ 10.1 \pm 17.6 $ \\
2458276.668550 & $	 -7.3 \pm 24.1 $ &  $   6.0 \pm 11.2 $ \\
2458278.682960 & $ -44.1 \pm 26.3 $ &  $ 19.6 \pm 17.1 $ \\
2458280.670353 & $  24.7 \pm 20.7 $ &  $  -7.9 \pm 12.7 $ \\
FIES & & \\
2458223.51513 & $  0.0 \pm  9.0 $ & $ 14.0 \pm  11.0 $ \\
2458233.50676 & $ -8.1 \pm  7.6 $ & $ 38.2 \pm  13.6 $ \\
\hline
\end{tabular}
\end{table}

\begin{table}
\centering
\caption{Relative RVs and BS variations for Qatar-9.}
\label{table:q9bRV}
\begin{tabular}{lrr}
\hline\hline
BJD$_{TDB}$       &RV (\ms)          &BS (\ms)       \\
\hline
2458172.753789 & $-508.1 \pm  62.8  $ & $   -29.4 \pm 33.6 $ \\
2458216.655360 & $ 100.7 \pm  90.8  $ & $    55.9 \pm 52.8 $ \\
2458259.747481 & $  -60.5 \pm  57.8  $ & $  -12.5 \pm 43.7 $ \\
2458266.688086 & $-550.1 \pm  37.1  $ & $-167.1 \pm 29.1 $ \\
2458273.684727 & $  -43.8 \pm  49.5  $ & $   28.0 \pm 34.0 $ \\
2458276.698788 & $	   0.0 \pm  57.8  $ & $   54.4 \pm 21.8 $ \\
2458277.715329 & $-213.2 \pm  74.9  $ & $   56.3 \pm 60.1 $ \\	
2458279.694742 & $	 38.7 \pm  49.4  $ & $    -9.1 \pm 43.0 $ \\
2458280.702564 & $-458.2 \pm  40.2  $ & $   23.6 \pm 29.8 $ \\

\hline
\end{tabular}
\end{table}

\begin{table}
\centering
\caption{Relative RVs and BS variations for Qatar-10.}
\label{table:q10bRV}
\begin{tabular}{lrr}
\hline\hline
BJD$_{TDB}$       &RV (\ms)          &BS (\ms)       \\
\hline
2458258.976713 & $   10.2  \pm 	36.9 $ & $  18.3 \pm 15.2 $ \\
2458259.824390 & $-179.3  \pm 	42.7 $ & $ -19.8 \pm 19.1 $ \\
2458263.950937 & $  -37.0	 \pm  28.7 $ & $ -31.0 \pm 19.0 $ \\
2458273.912527 & $     0.0	 \pm  34.9 $ & $ -29.1 \pm   9.0 $ \\
2458274.928598 & $-175.8	 \pm  34.9 $ & $ -28.0 \pm 15.8 $ \\
2458277.931350 & $-216.8	 \pm  39.2 $ & $ -20.9 \pm 19.8 $ \\
2458278.755400 & $   57.9	 \pm  39.3 $ & $  38.5 \pm 30.2 $ \\
2458280.931487 & $-170.8	 \pm  30.0 $ & $ -24.8 \pm 15.4 $ \\
2458281.917044 & $   55.7	 \pm  23.3 $ & $    6.6 \pm 15.9 $ \\
2458292.804140 & $-204.4	 \pm  34.8 $ & $   -3.7 \pm 11.9 $ \\
2458296.861241 & $ 152.0	 \pm  38.2 $ & $  64.4 \pm 31.0 $ \\
2458300.930706 & $-232.4	 \pm  33.5 $ & $  15.3 \pm 16.6 $ \\
2458301.790512 & $   39.0	 \pm  33.1 $ & $  14.8 \pm 16.6 $ \\
2458386.632867 & $-174.7  \pm  30.1 $ & $  40.9 \pm 21.0 $ \\
2458387.655367 & $  -13.7  \pm  32.6 $ & $ -10.0 \pm 18.6 $ \\
2458389.611784 & $-130.1  \pm  37.6 $ & $ -49.1 \pm 20.5 $ \\
\hline
\end{tabular}
\end{table}

\begin{figure}
\centering
\includegraphics[width=8.5cm]{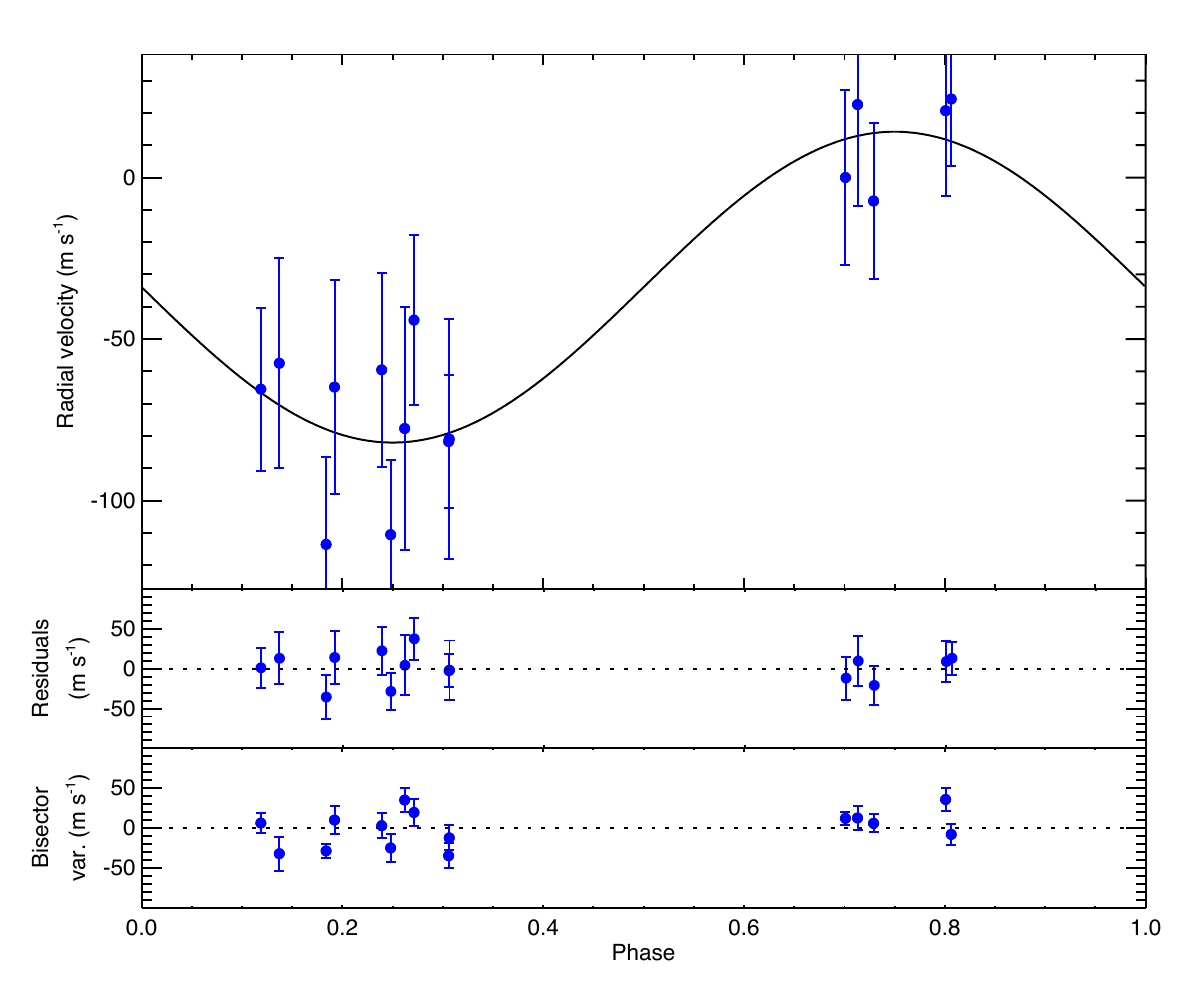}
\caption{Orbital solution for Qatar-8b, showing the velocity curve and observed 
velocities and the bisector values.}
\label{figure:q8bRV}
\end{figure}

\begin{figure}
\centering
\includegraphics[width=8.5cm]{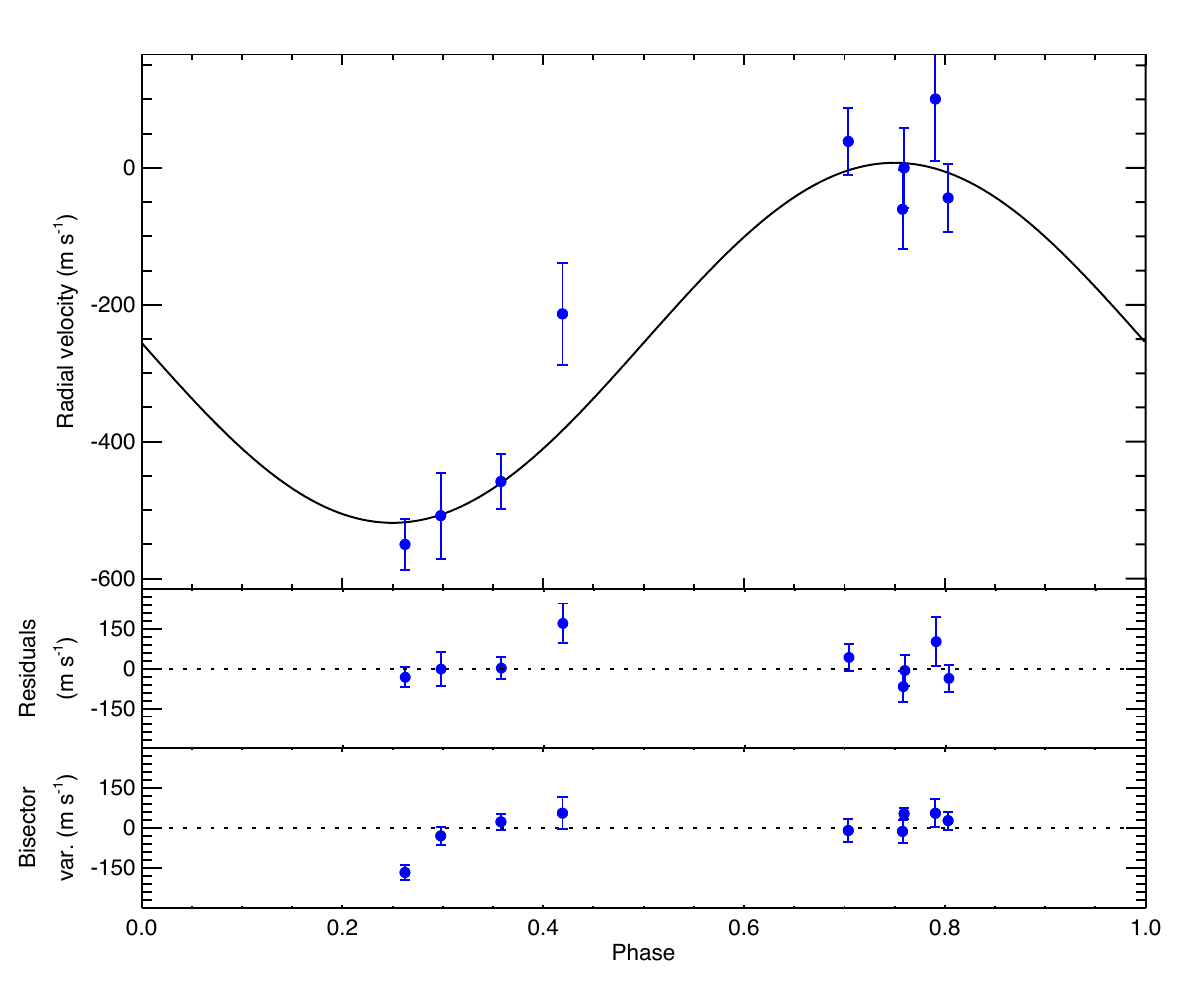}
\caption{Orbital solution for Qatar-9b, showing the velocity curve and observed 
velocities and the bisector values.}
\label{figure:q9bRV}
\end{figure}

\begin{figure}
\centering
\includegraphics[width=8.5cm]{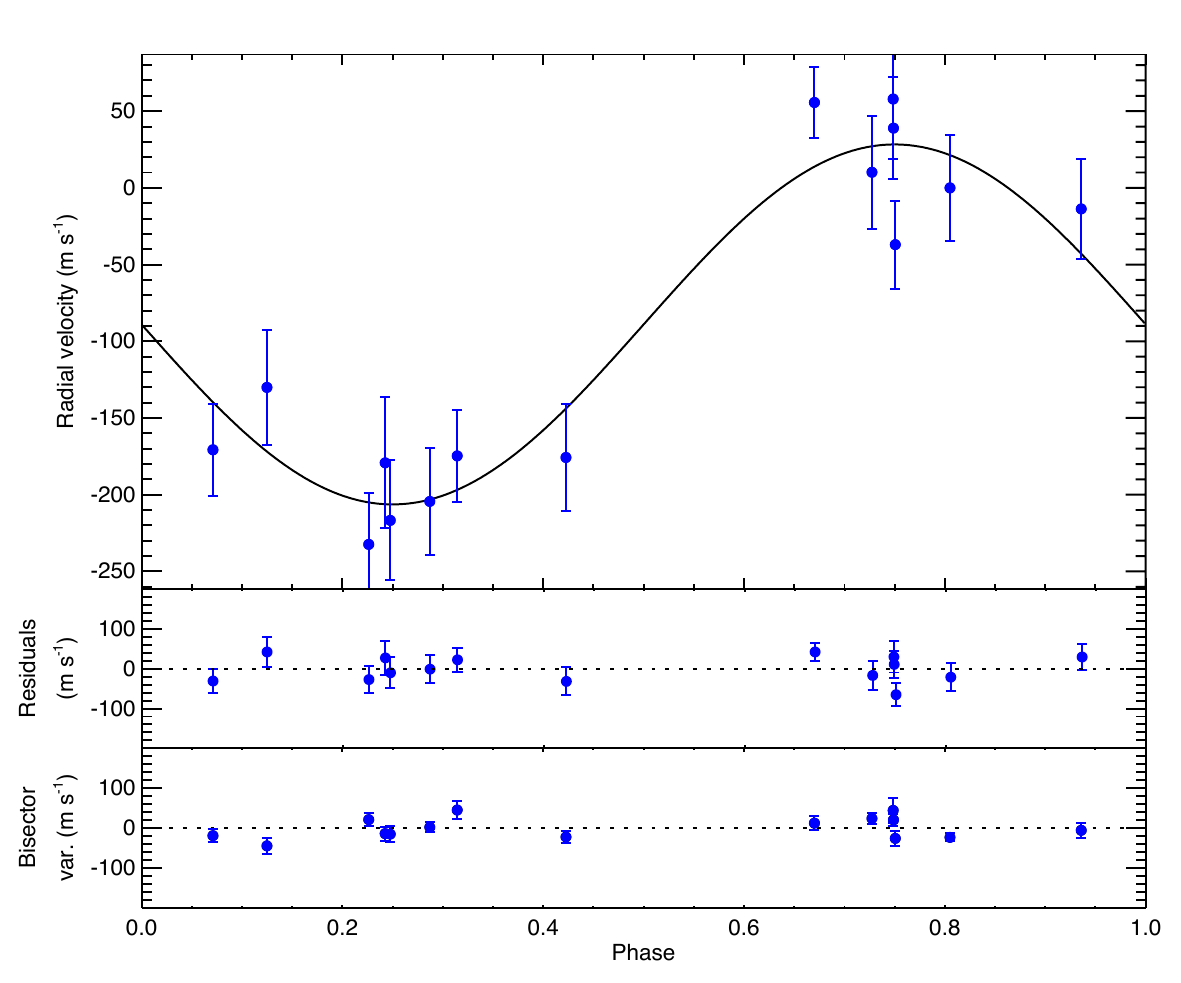}
\caption{Orbital solution for Qatar-10b, showing the velocity curve and observed 
velocities and the bisector values.}
\label{figure:q10bRV}
\end{figure}

For each one of the three target stars we obtained the following TRES spectra: 
(a) Qatar-8 -- 15 spectra between March 6, 2017 -- June 11, 2018 with exposure 
times in the range 10--30 min and an average signal-to-noise ratio per resolution 
element (SNRe) $\sim$34 at the peak of the continuum in the echelle order 
centered on the Mg b triplet near 519 nm; (b) Qatar-9 -- 9 spectra between 
February 23 -- June 11, 2018 all with individual exposure times 60 min and an 
average SNRe $\sim$14.9; (c) Qatar-10 -- 16 spectra between May 20 -- September 28, 
2018, with exposure times in the range 20--40 min and an average SNRe $\sim$30. 
For each of our target stars, we cross-correlated each observed spectrum against 
the ''reference'' spectrum (taken to be the strongest among the spectra of the given 
star) to obtain relative radial velocities (RVs). This was done for a set of echelle 
orders (in an order-by-order fashion) chosen so as to have both good SNRe and 
minimal telluric line contamination.These RVs are reported in Tables \ref{table:q8bRV}, 
\ref{table:q9bRV} and \ref{table:q10bRV} (with the time stamps in Barycentric Julian 
Date in Barycentric Dynamical time, BJD$_{\rm TDB}$) and plotted in Figures 
\ref{figure:q8bRV}, \ref{figure:q9bRV} and \ref{figure:q10bRV}. The observation that 
was used for the template spectrum has, by definition, a RV of 0.0 \kms, and the 
error on the template RV is the median of the uncertainties of all the orders. In order 
to be sure that the periodic RV signal detected in all three cases is due to orbital 
motion and exclude other astrophysical phenomena that could potentially produce a 
similar periodic signal, we also derived the line profile bisector span values (BS, 
lower panel in Figures \ref{figure:q8bRV}, \ref{figure:q9bRV} and \ref{figure:q10bRV}).  
\citet{buchhave2010} describe the procedures used above for both the RV and the 
BS measurements in more detail.
The absolute center-of-mass velocity of each system is determined in two steps as 
the precise RV of each spectrum is measured against the strongest observed 
spectrum of the same star. That reference spectrum could be anywhere on the RV 
curve, and so a relative systemic velocity ($\gamma_{\rm rel}$) is determined when 
fitting the Keplerian model. The absolute systemic velocity is then the sum of 
$\gamma_{\rm rel}$ and the absolute RV offset of the reference spectrum which is 
determined by cross-correlating the Mg b order of the respective reference spectrum 
against the CfA library of synthetic templates. We also correct by $-0.61$ \kms, 
because the CfA library does not include the gravitational redshift. This offset has 
been determined empirically by many observations of IAU Radial Velocity Standard 
Stars. We quote an uncertainty of $\pm 0.1$ \kms\ in the resulting absolute velocity, 
which is an estimate of the residual systematic errors in the IAU Radial Velocity 
Standard Star system. Note that the error in determining the absolute center of mass 
RV of each system does not affect the determination of the planetary parameters. 
Although a dedicated paper on the assessment of the TRES absolute zero point has 
not been published, we refer the interested reader to \cite{Quinn2014, Quinn2015}, 
which gives a detailed account of the instrument RV precision, stability, sources of 
error and steps used to bring observed stellar velocities to the absolute scale of the 
IAU radial velocity system.

For Qatar-8 we also obtained 2 RV measurements, on April 15 and 25, 2018, with 
the high-resolution Fiber-fed Echelle Spectrograph (FIES, \citealt{Telting2014}) on 
the 2.5\,m Nordic Optical Telescope (NOT) at the Observatorio del Roque de los 
Muchachos (ORM) on the island of La Palma, Canary Islands, Spain. We used 
FIES in its high resolution mode $R \sim$ 67,000 and a velocity resolution 
element of 4.8 \kms\ FWHM. For the first observation we obtained a single 30 min 
exposure spectrum, while for the second observation we obtained three consecutive 
15 min exposure spectra. Similar to the TRES observations, the wavelength 
calibration was established using exposures of a Th-Ar lamp illuminating the science 
fiber bracketing the target exposure spectra. Relative RV measurements were 
obtained through the cross-corellation technique described above and using the 
spectrum obtained on April 15, 2018 ($T_{\rm exp} = 30$ min) as a template. Each 
one of the 3 spectra obtained on April 25, 2018 was measured separately against 
the template and the three measurements were averaged. 

\section{Analysis and Results} \label{sec:Analysis} 

\subsection{Rejecting False Positives} \label{subsec:FPP}

For all three targets, the observed transit light curves combined with the shape and 
amplitude of the RV variations, are well described by a planet orbiting a single star. 
Nevertheless, there are astrophysical scenarios not involving a planet that could mimic 
such a behaviour. These include an eclipsing binary -- either a background or in a 
hierarchieal tripple system -- blended with the primary much brighter star. Next we put 
forward arguments that allow us to exclude such scenarios. 

It has been well established (e.g., \citealt{Queloz01}, \citealt{Torres05}) that if the observed 
RV pattern were a result of a blend with an eclipsing binary, the spectral line bisectors (see 
Tables \ref{table:q8bRV}, \ref{table:q9bRV}, and \ref{table:q10bRV}) would follow a similar 
pattern, that is, they would vary in phase with the photometric period and with a similar 
amplitude. The measured line bisectors are shown in the bottom panel in Figures 
\ref{figure:q8bRV}, \ref{figure:q9bRV}, and \ref{figure:q10bRV} on the same scale as 
the RV residuals from the Keplerian orbit fit. No obvious pattern is seen in all three cases. 
To quantify that we performed a significance test on the Pearson's correlation between the 
BS and RV for each star, and also the BS and the Keplerian fit model. In each of the three 
cases, the correlation between the BS and the RV (or between the BS and the Keplerian 
model) is insignificant, while the correlation between the RV and the Keplerian model is 
highly significant. Numerically, the probability of chance RS/RV correlation is higher than 
the typical significance level (5\%): Qatar-8b --  $p$-values, $p = 0.061$; Qatar-9b -- 
$p = 0.073$; Qatar-10b -- $p = 0.067$. This supports our argument that the observed 
RV pattern is a result of a gravitationally induced motion from a planet orbiting a single star.

A further argument in favor of the planet scenario can be drawn from the equal transit 
depths across all filters (accounting for limb darkening effects). We do, however, note 
that an eclipse of a stellar companion with very similar colors would also lead to equal 
depths at different wavelengths. As such, we consider the equal depths as supportive 
argument only, albeit in full agreement with our conclusion regarding the planetary nature 
of the transits.


\subsection{Planetary System Parameters} \label{subsec:GlobalFit}

\begin{figure}
\centering
\includegraphics[width=5.9cm]{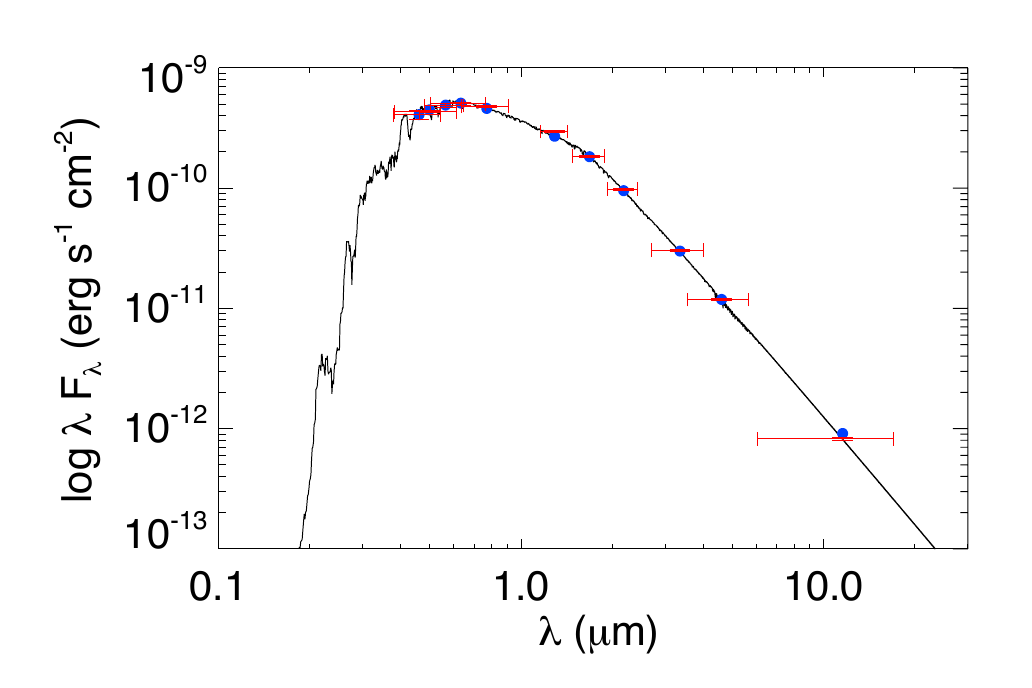}
\includegraphics[width=5.9cm]{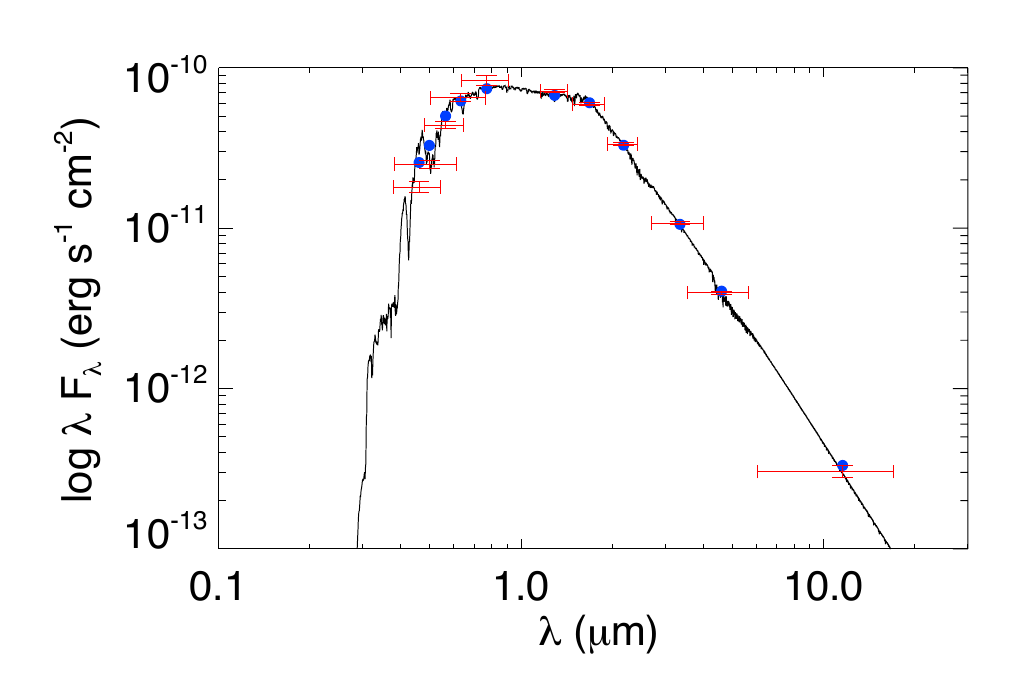}
\includegraphics[width=5.9cm]{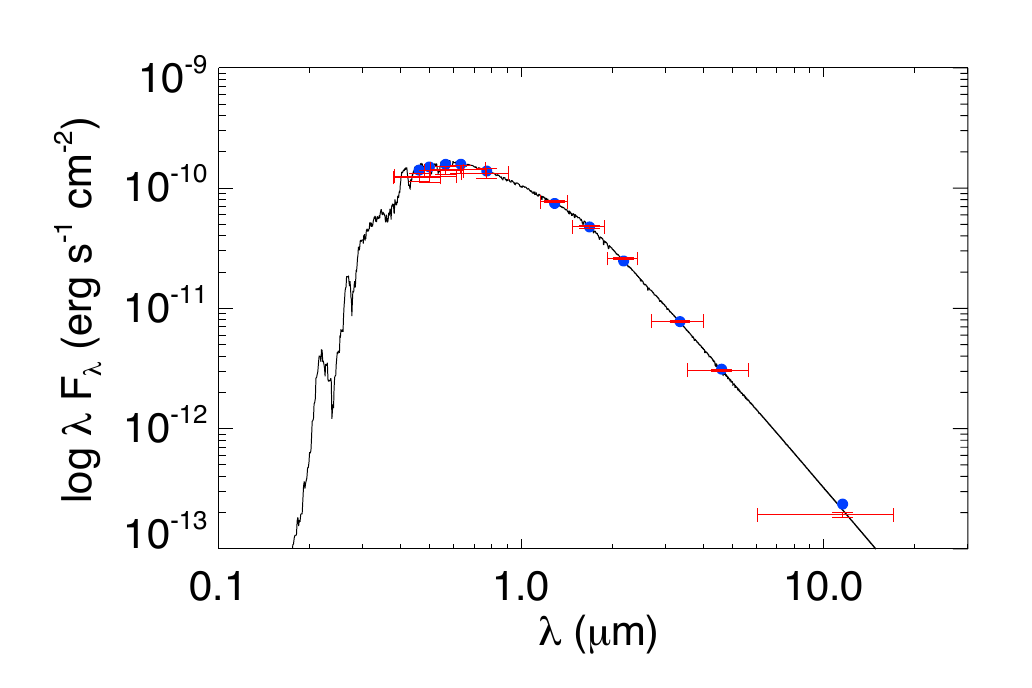}
\caption{SED fit for the hosts stars of Qatar-8b (left panel), Qatar-9b (middle panel), 
and Qatar-10b (right panel). Photometric measurements used in each fit (Table 
\ref{table:StarPars}) are plotted here as error bars, where the vertical bars represent 
the quoted 1$\sigma$ measurement uncertainties, and the horizontal bars mark the 
effective width of the passbands. The solid curve is the best fit SED from the 
NextGen library of models from the global EXOFASTv2 fit.}
\label{fig:SEDfit}
\end{figure}


\begin{table*}
\centering
\caption{Basic observational and spectroscopic parameters of Qatar-8b, 9b, 10b host stars 
and photometry used for the SED fit.}
\label{table:StarPars}
\begin{tabular}{lllllll}
\hline \hline
Parameter & Description & \multicolumn{3}{c}{Value}  & Ref. \\ 
\hline
\multicolumn{2}{l}{Names}	& Qatar-8b & Qatar-9b & Qatar-10b & \\
	&3UC	& 322-045544	& 302-100935	& 320-060170 & \\
	&2MASS	& J23540364+3701185	& J23540364+3701185	& J23540364+3701185 & \\

 \multicolumn{2}{l}{Astrometry} & & & \\
$\alpha_{\mathrm{2000}}$ & RA (J2000) & 
	$10^{\mathrm{h}}29^{\mathrm{m}}38.962^{\mathrm{s}}$ & 
	$10^{\mathrm{h}}42^{\mathrm{m}}59.543^{\mathrm{s}}$ & 	
	$18^{\mathrm{h}}57^{\mathrm{m}}46.537^{\mathrm{s}}$ & 1 \\    
$\delta_{\mathrm{2000}}$  & DEC (J2000) & 
	$+70^{\mathrm{o}}31^{\prime}37.50^{\prime\prime}$ & 
	$+60^{\mathrm{o}}57^{\prime}50.83^{\prime\prime}$ &
	$+69^{\mathrm{o}}34^{\prime}15.01^{\prime\prime}$ & 1 \\

\multicolumn{2}{l}{Photometry} & & & \\
$B$	& Johnson $B$, mag	& 12.132 $\pm$ 0.084 & 15.515 $\pm$ 0.088 & 13.437 $\pm$ 0.080 & 2 \\       
$V$	& Johnson $V$, mag	& 11.526 $\pm$ 0.046 & 14.133 $\pm$ 0.048 & 12.879 $\pm$ 0.079 & 2 \\       
$g$	& Sloan $g$, mag	& 11.806 $\pm$ 0.018 & 14.894 $\pm$ 0.065 & 13.147 $\pm$ 0.124 & 2 \\       
$r$	& Sloan $r$, mag	& 11.355 $\pm$ 0.028 & 13.569 $\pm$ 0.061 & 12.712 $\pm$ 0.079 & 2 \\    
$i$	& Sloan $i$, mag	& 11.174 $\pm$ 0.015 & 13.074 $\pm$ 0.078 & 12.569 $\pm$ 0.109 & 2 \\     
$J$ 	& 2MASS $J$, mag	& 10.299 $\pm$ 0.023 & 11.835 $\pm$ 0.019 & 11.754 $\pm$ 0.022 & 3 \\
$H$	& 2MASS $H$, mag	& 10.005 $\pm$ 0.024 & 11.235 $\pm$ 0.017 & 11.474 $\pm$ 0.023 & 3 \\
$K$	& 2MASS $K$, mag	&   9.937 $\pm$ 0.015 & 11.104 $\pm$ 0.023 & 11.382 $\pm$ 0.017 & 3 \\
W1	& WISE1, mag     	&   9.890 $\pm$ 0.023 & 11.021 $\pm$ 0.024 & 11.375 $\pm$ 0.023 & 4 \\              
W2	& WISE2, mag    	&   9.932 $\pm$ 0.021 & 11.120 $\pm$ 0.021 & 11.413 $\pm$ 0.020 & 4\\                
W3	& WISE3, mag    	&   9.902 $\pm$ 0.033 & 10.986 $\pm$ 0.090 & 11.348 $\pm$ 0.067 & 4 \\               

\multicolumn{2}{l}{Spectroscopic parameters} & & & & & \\
       & Spectral type & G0V & K5V & F7V & this work & \\
\teff  & Effective temperature, K 	& 5687$\pm$50 		& 4363$\pm$51   & 6123$\pm$50   & this work \\
\logg & Gravity, cgs                	& 4.22$\pm$0.10 	& 4.65$\pm$0.10 & 4.36$\pm$0.10 & this work \\
\mh   & Metallicity                    	& 0.0$\pm$0.08   	& 0.25$\pm$0.08 & 0.40$\pm$0.08 & this work \\
$\gamma_{\rm abs}$ &Systemic velocity, \kms\ & $5.57\pm0.10$ & $2.92\pm0.10$ & $-25.49\pm0.10$ & this work \\
$v_{\rm rot}$ & Rotational velocity, \kms & 2.7$\pm$0.5 & 4.3$\pm$0.5 & 5.9$\pm$0.5 & this work \\
\hline
\end{tabular}
\tablerefs{(1) GAIA DR2 \url{http://gea.esac.esa.int/archive/}, \\
	(2) APASS9 \url{http://www.aavso.org/apass}, \\
	(3) 2MASS \url{http://irsa.ipac.caltech.edu/Missions/2mass.html}, \\
	(4) WISE \url{http://irsa.ipac.caltech.edu/Missions/wise.html} \\
	}
\end{table*}

Physical properties of each system were determined through a global model fit using the 
{\sc EXOFASTv2} package. A detailed description of {\sc EXOFASTv2} can be found in 
\cite{Eastman2017} and \cite{Rodrigues2017}. For each of the 
three exoplanetary systems the global fit includes the RV measurements listed in Tables 
\ref{table:q8bRV}, \ref{table:q9bRV} and \ref{table:q10bRV}, the follow-up transit light 
curves shown in Figures \ref{fig:q8bTR}, \ref{fig:q9bTR}, and \ref{fig:q10bTR}, respectively, 
the distance and boradband photometry for each star, and the uses priors on the stellar 
atmospheric parameters (\teff\ and \feh) determined from the available specta. 

Effective temperature (\teff), surface gravity (\logg), metallicity (\mh) and projected 
rotational velocity ($v_{\rm rot}$) -- for the host stars were determined using the Stellar 
Parameter Classification tool (SPC, \citealt{SPC}). SPC works by cross correlating an 
observed spectram against a grid of synthetic spectra based on the Kurucz atmospheric 
models. We used the ATLAS9 grid of models with the Opacity Distribution Functions 
from \cite{ODF}. In Table \ref{table:StarPars} we list the weighted mean values and 
the associated uncertainties of the stellar athmospheric parameters determined from 
the SPC analysis of each individual spectrum. 

Broadband photometric surveys provide measurements across the electromagnetic 
spectrum for all three host stars, from the optical (APASS) to the mid-IR (WISE). 
These measurements are gathered in Table\,\ref{table:StarPars} and used to fit a 
model Spectral Energy Distribution (SED) for each one of the stars as described below. 
The resulting SED fits are shown in Figure\,\ref{fig:SEDfit}.
 
In the global fit we apply Gaussian priors on the parallax from {\it Gaia} DR2, including 
the offset determined by \citet{Stassun2016} and impose an upper limit on the $V$-band 
extinction from the Galactic dust reddening maps (\citealt{Extinction}). The limb darkening 
coefficients (LDCs) were fit with a prior derived from an interpolation from the 
\cite{LDcoeff} tables for each band. We note that while all transit light curves were 
normalized as described in \ref{subsec:FollowPhot} in the global fit {\sc EXOFASTv2}
treats the baseline flux (the flux outside the transit) as a free constant and adjusts its 
value to unity. Because all transit light curves were normalized in advance this adjustment 
is naturally very small. 

All of the host stars properties were determined during the global fit. {\sc EXOFASTv2} 
simultaneously uses the SED, the stellar density and limb darkening constraint from the 
transit, the MIST isochrones (\citealt{MESA}, \citealt{MIST}), priors from SPC (\teff, \feh) 
and {\it Gaia} (parallax), and an upper limit on the reddening, to simultaneously determine 
all the stellar properties. The stellar radius is predominantly constrained by the SED and 
{\it Gaia} parallax, while the stellar age and mass are predominantly constrained by the 
MIST isochrones and spectroscopic priors. In addition, consistency between the stellar 
mass and radius derived from these methods and the stellar density from the transit is 
strictly required.

In fitting Qatar-8b, 9b, and 10b we only considered circular orbits and kept the eccentricity 
fixed to zero. On one hand, our RV data is not of high enough quality to allow investigation 
of potential small departures from circularity, and on the other hand we expect the planets 
orbits to have circularized. Following the equations from \cite{Jackson08}, and using the 
values from Table \ref{table:GlobalFit} of $\mstar$, $\rstar$, $\mpl$, $\rpl$, and $a/\rstar$ 
we estimate orbit circularization time-scales $\tau_{\rm circ} \sim 0.04, 0.01, 0.01$ Gyr, 
respectively, for the entire range of tidal quality factors $Q_{\star}$ and $Q_{\rm P}$ 
considered by the authors ($10^{4}-10^{8}$ for each $Q$). This is much lower than the 
estimated age of the host stars and, thus, we expect the planet orbit to have circularized.

A precise orbital period for each system is also obtained during the global fit. The best 
ephemeris for each star is calculated by fitting all transits simultaneously to a linear 
ephemeris within {\sc EXOFASTv2}: 

\begin{equation}
T_{\rm C} = 2458210.83980(85) + 3.71495(100)\,E
\label{eq:OrbEphemerisQ8b}
\end{equation}

\begin{equation}
T_{\rm C} = 2458227.75643(27) + 1.540731(38)\,E 
\label{eq:OrbEphemerisQ9b}
\end{equation}

\begin{equation}
T_{\rm C} = 2458247.90746(36) + 1.645321(10)\,E 
\label{eq:OrbEphemerisQ10b}
\end{equation}

\noindent 
where $E$ is the number of cycles after the reference epoch, which we take to be the 
transit time that minimizes the covariance between $T_{\rm C}$ and period, and the 
numbers in parenthesis denote the uncertainty in the last two digits. Equations 
\ref{eq:OrbEphemerisQ8b}, \ref{eq:OrbEphemerisQ9b}, and \ref{eq:OrbEphemerisQ10b} 
correspond to Qatar-8b, 9b and 10b, respectively. 

Table \ref{table:GlobalFit} summarizes the physical parameters of each planetary system. 
We note, that for Qatar-8b, for which we have RV measurements from two different 
telescopes and instruments, {\sc EXOFASTv2} fits the relative offset for each RV data 
set separately. These are the reported $\gamma_{\rm rel}$ values in the table.
The Safronov number is not used in the current paper and is provided in Table 
\ref{table:GlobalFit} for completeness, as it may be useful for other studies. 

\begin{table*}
\centering
\caption{Median values and 68\% confidence intervals.}
\label{table:GlobalFit}
\begin{tabular}{lcccc}
\hline
Parameter & Units & Qatar-8b & Qatar-9b & Qatar-10b \\
\hline
\multicolumn{2}{l}{Stellar Parameters:} &  \\
    ~~~$M_{*}$\dotfill	   &Mass (\msun)\dotfill  		& $1.029\pm0.051$	& $0.719\pm0.024$	& $1.156\pm0.068$ \\
    ~~~$R_{*}$\dotfill	   &Radius (\rsun)\dotfill  		& $1.315\pm0.020$	& $0.696\pm0.008$	& $1.254\pm0.026$ \\
    ~~~$L_{*}$\dotfill	   &Luminosity (\lsun)\dotfill 		& $1.690\pm0.068$	& $0.151\pm0.004$ 	& $1.993\pm0.094$ \\
    ~~~$\rho_*$\dotfill	   &Density (g/cm$^{3}$)\dotfill 	& $0.641\pm0.024$	& $3.015\pm0.086$ 	& $0.823\pm0.062$ \\
    ~~~$\log(g_*)$\dotfill &Surface gravity (cgs)\dotfill 	& $4.214\pm0.016$	& $4.610\pm0.010$ 	& $4.303\pm0.027$ \\
    ~~~$\teff$\dotfill	&Effective temperature (K)\dotfill	& $5738\pm51$  	& $4309\pm31$    	& $6124\pm46$ \\
    ~~~$\feh$\dotfill	   &Metallicity\dotfill         		& $0.025\pm0.071$	& $0.252\pm0.076$ 	& $0.016\pm0.089$ \\
    ~~~$\tau_{\rm MIST}$\dotfill &Age (Gyr)\dotfill 		& $8.3\pm2.1$    	& $7.5\pm4.5$   		& $3.2\pm1.9$  \\
    ~~~$A_{V}$\dotfill             & Extinction (mag)\dotfill	& $0.063\pm0.042$ 	& $0.016\pm0.010$ 	& $0.119\pm0.062$ \\
    ~~~$\pi$\dotfill  &Parallax (mas)\dotfill           		& $3.614\pm0.043$    & $4.730\pm0.036$	& $1.855\pm0.035$ \\
    ~~~$d$\dotfill  &Distance (pc)\dotfill           		& $276.7\pm3.4$    	& $211.4\pm1.6$      	& $539\pm10$  \\
\multicolumn{2}{l}{Planetary Parameters:} &  \\
    ~~~$P$\dotfill   &Period (days)\dotfill 	& $3.71495\pm0.00100$ & $1.540731\pm0.000038$ & $1.645321\pm0.000010$ \\
    ~~~$a$\dotfill   &Semi-major axis (AU)\dotfill       		& $0.0474\pm0.0008$ & $0.0234\pm0.0003$ & $0.0286\pm0.0006$ \\
    ~~~$M_{P}$\dotfill &Mass (\mj)\dotfill              			& $0.371\pm0.062$	 & $1.19\pm0.16$ 	  & $0.736\pm0.090$ \\
    ~~~$R_{P}$\dotfill &Radius (\rj)\dotfill                			& $1.285\pm0.022$	 & $1.009\pm0.014$ 	  & $1.543\pm0.040$ \\
    ~~~$\rho_{P}$\dotfill &Density (g/cm$^{3}$)\dotfill 		& $0.216\pm0.037$ 	 & $1.43\pm0.20$ 	  & $0.248\pm0.036$ \\
    ~~~$\log(g_{P})$\dotfill &Surface gravity\dotfill 			& $2.745\pm0.080$ 	 & $3.460\pm0.063$ 	  & $2.884\pm0.059$ \\
    ~~~$T_{eq}$\dotfill &Equilibrium Temperature (K)\dotfill 	& $1457\pm14$   	 & $1134\pm9$    	  & $1955\pm25$ \\
    ~~~$\Theta$\dotfill &Safronov Number\dotfill           		& $0.0265\pm0.0044$ & $0.0764\pm0.0100$ & $0.0236\pm0.0028$ \\
\multicolumn{2}{l}{RV Parameters:} &  \\
    ~~~$K$\dotfill &RV semi-amplitude (m/s)\dotfill           			& $47.7\pm8.0$    	& $259\pm35$     	& $114\pm13$ \\
    ~~~$\gamma_{\rm rel}$\dotfill &Relative RV offest, TRES (m/s)\dotfill 	& $-34.2\pm6.8$ 	& $-252\pm32$ 		& $-89\pm11$ \\
    ~~~$\gamma_{\rm rel}$\dotfill &Relative RV offest, FIES (m/s)\dotfill 	& $25\pm13$      	&                         	&  \\
    ~~~$e$\dotfill &Eccentricity (fixed)\dotfill               				& $0$                    	& $0$                   	& $0$ \\
\multicolumn{2}{l}{Primary Transit Parameters:} &  \\
    ~~~$T_C$\dotfill &Time of transit (\bjdtdb)\dotfill & $2458210.83980\pm0.00085$ & $2458227.75643\pm0.00027$ & $2458247.90746\pm0.00036$ \\
    ~~~$R_{P}/R_{\star}$\dotfill &Radius of planet in stellar radii\dotfill	& $0.1005\pm0.0008$ & $0.1489\pm0.0009$ & $0.1265\pm0.0010$ \\
    ~~~$a/R_{\star}$\dotfill &Semi-major axis in stellar radii\dotfill 	   	& $7.761\pm0.100$	 & $7.236\pm0.069$ 	  & $4.90\pm0.12$\\
    ~~~$i$\dotfill &Inclination (degrees)\dotfill                                        	& $89.29\pm0.70$ 	 & $89.23\pm0.64$ 	  & $85.87\pm0.96$\\
    ~~~$b$\dotfill &Impact Parameter\dotfill                                           	& $0.096\pm0.093$	 & $0.097\pm0.079$ 	  & $0.379\pm0.055$ \\
    ~~~$T_{14}$\dotfill &Total duration (days)\dotfill           	& $0.1678\pm0.0017$  & $0.0778\pm0.0004$    & $0.1155\pm0.0009$ \\
\hline
\end{tabular}
\end{table*}
\noindent

\section{Discussion and Conclusions} \label{sec:conclusions}
\begin{figure}
\centering
\includegraphics[width=8.8cm]{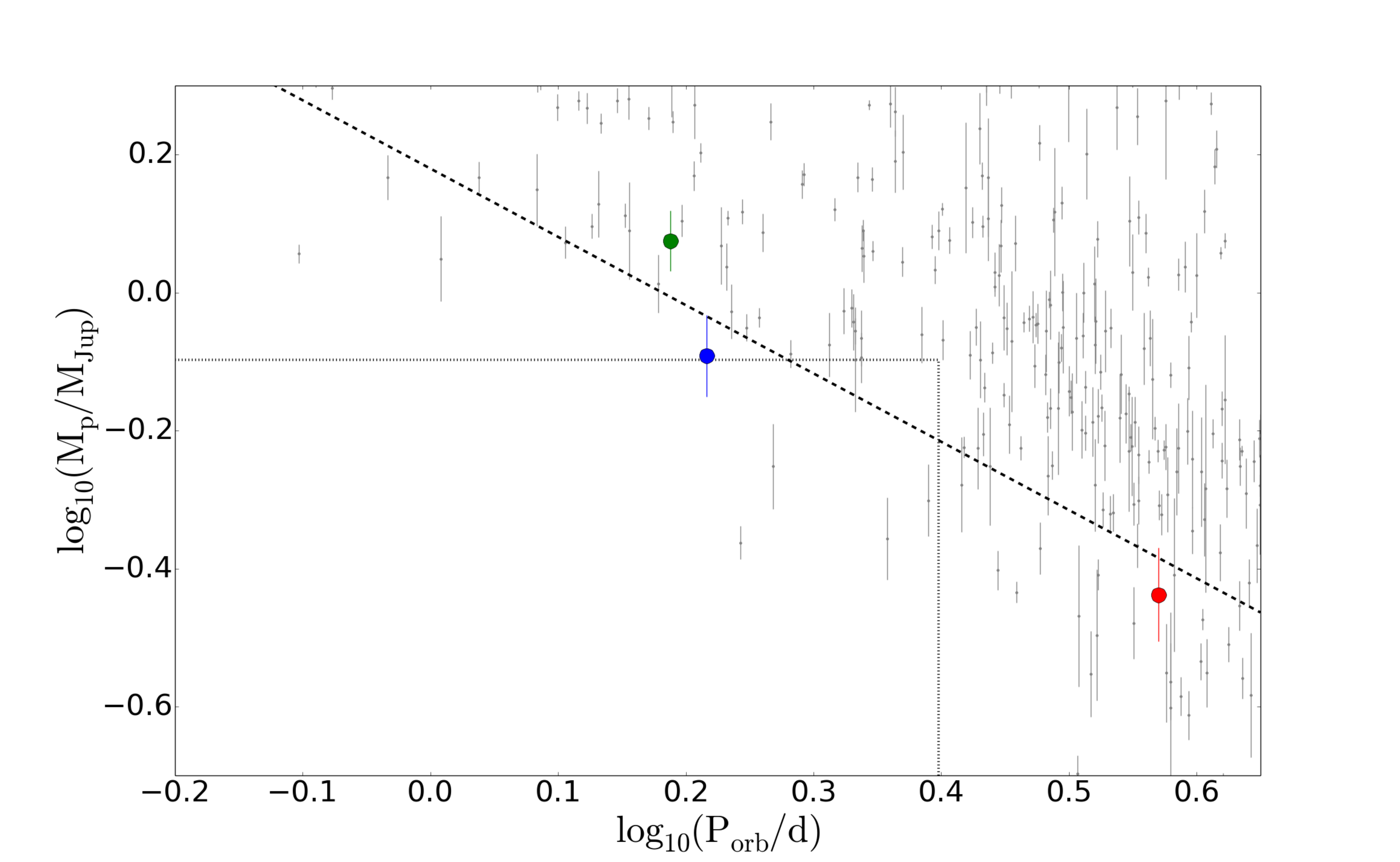}
\caption{Orbital period versus planet mass. The planets data (gray points) are from 
TEPCat, while Qatar-8b, 9b and 10b are plotted as the red, green and blue points, respectively. 
The dotted box is the sub-Jupiter desert as defined by \citet{desert1}, while the dashed line is 
the upper limit of the same, as defined by \citet{desert2}.}
\label{fig:porb_vs_mass}
\end{figure}

\begin{figure}
\centering
\includegraphics[width=8.8cm]{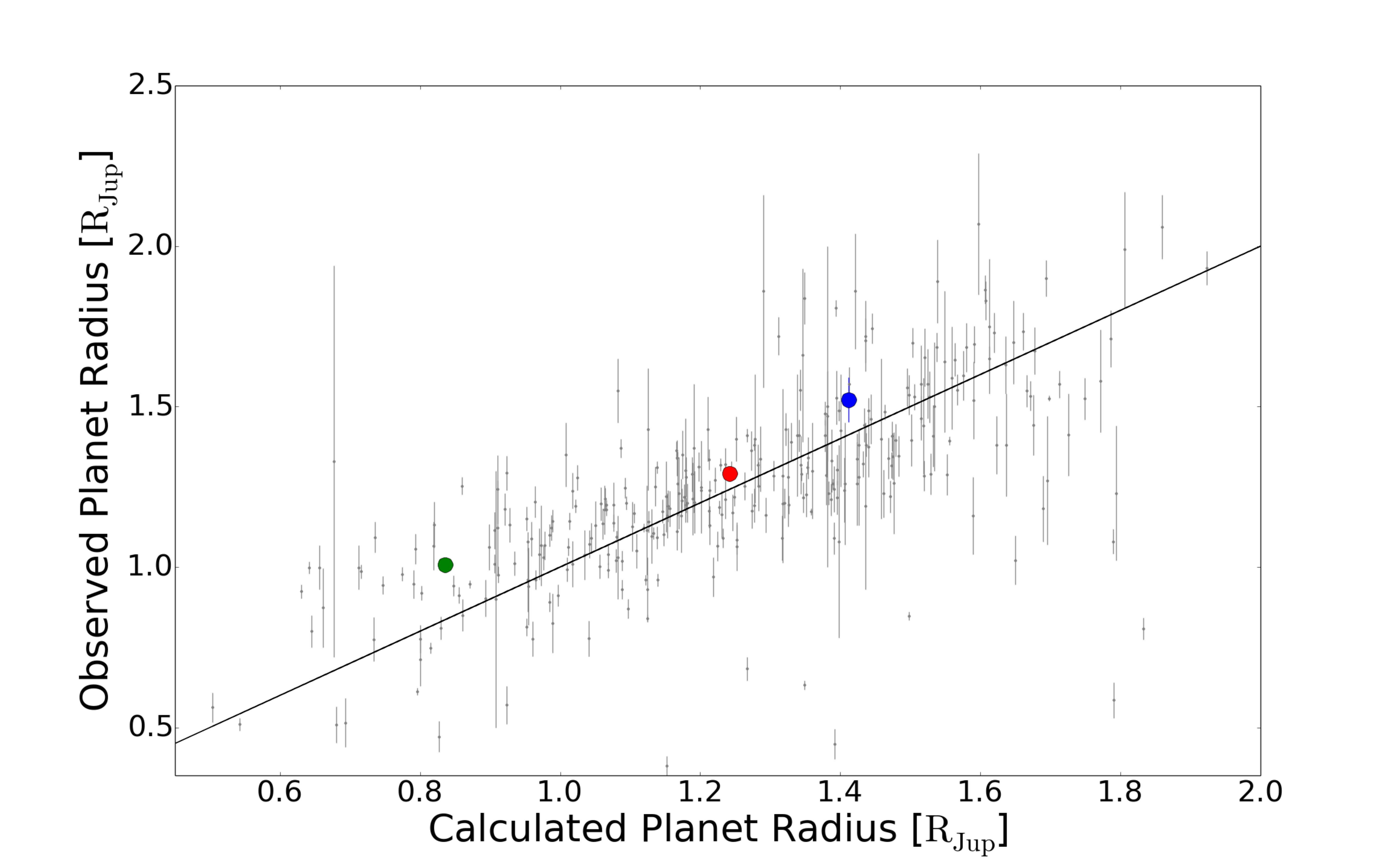}
\includegraphics[width=8.8cm]{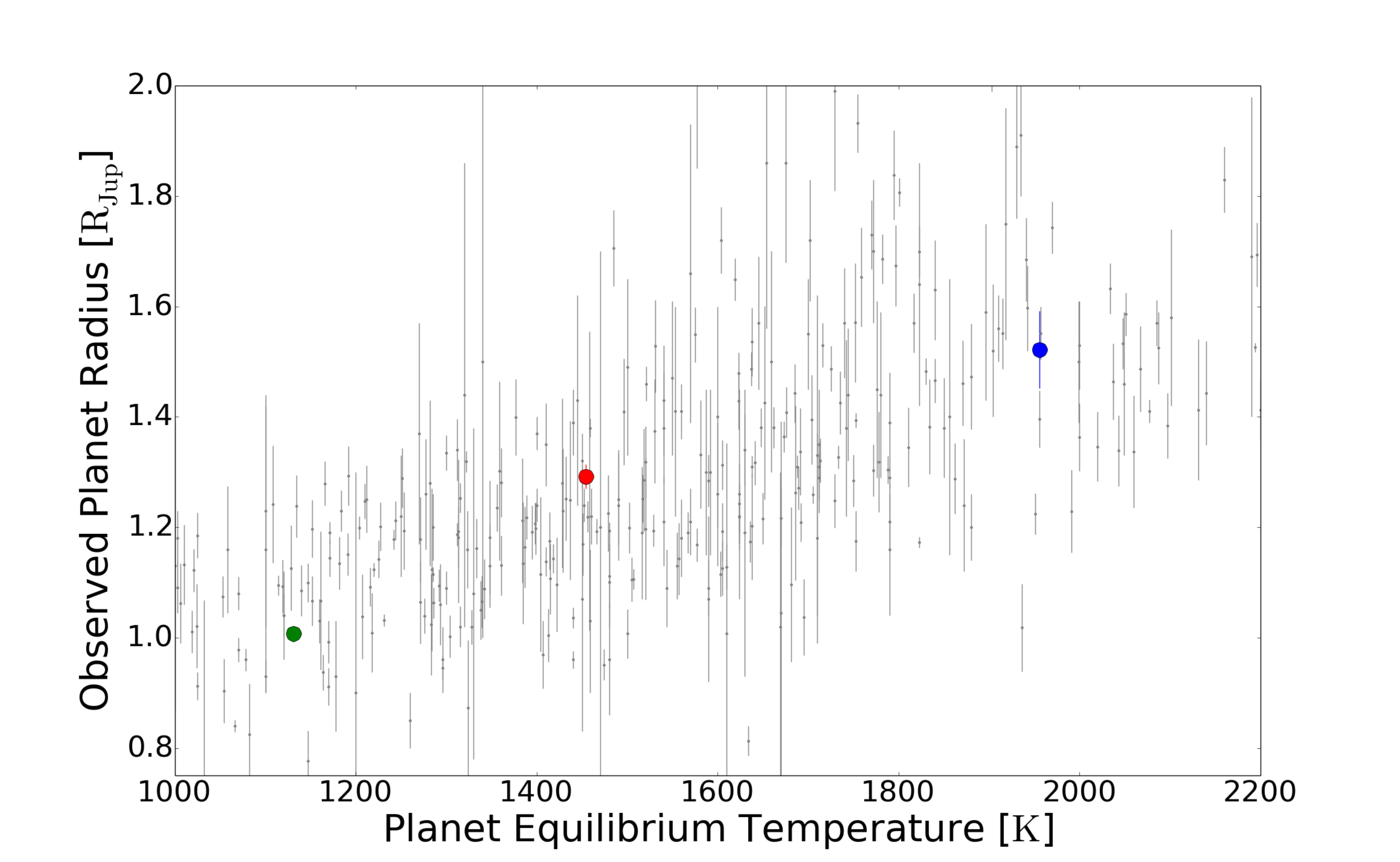}
\caption{{\it Left panel:} theoretical planetary radii, calculated from \citet{Enoch2012}, 
versus observed planetary radii. {\it Right panel:} planetary equilibrium temperatures 
versus observed planetary radii. As in Figure\,\ref{fig:porb_vs_mass}, in both panels, 
gray points are data from TEPCat, while the R-G-B dots represent Qatar-8b, 9b and 
10b, respectively.}
\label{fig:cror_tor}
\end{figure}

In this paper we present Qatar-8b, 9b and 10b, a transiting hot Saturn and two transiting 
hot Jupiters identified by QES. We combine follow-up photometric and spectroscopic 
observations, together with available broadband photometry and GAIA measurements, 
to calculate a full set of physical parameters for the planets and their host stars. In Figures 
\ref{fig:porb_vs_mass} and \ref{fig:cror_tor} we present the new discoveries in the context 
of the current state of the field. To produce these Figures, we made use of the well-studied 
sample of planets from the Transiting Extrasolar Planets Catalogue \citep[TEPCat;][online 
version\footnote{http://www.astro.keele.uk/jkt/tepcat} as of October 10, 2018]{tepcat}

Qatar-8b is a typical example of a hot Saturn, Qatar-9b is slightly more massive than 
Jupiter itself, but with a similar radius, and Qatar-10b is a sub-Jupiter mass planet, very 
similar to HATS-9b \citep{HATS9b} and WASP-142b \citep{WASP142b}. In the left panel 
of Figure \ref{fig:cror_tor} we present a predicted vs.\ observed planetary radii plot based 
on the widely used \citet{Enoch2012} relations. At face value, the \citet{Enoch2012} relations 
underpredict the observed radius of Qatar-9b, while the observed radii of Qatar-8b and 
Qatar-10b are close to the theoretical predictions. We note however, that none of the three 
new exoplanets stand out from the rest and follow the general trend well. 



\begin{figure}
\centering
\includegraphics[width=8.8cm]{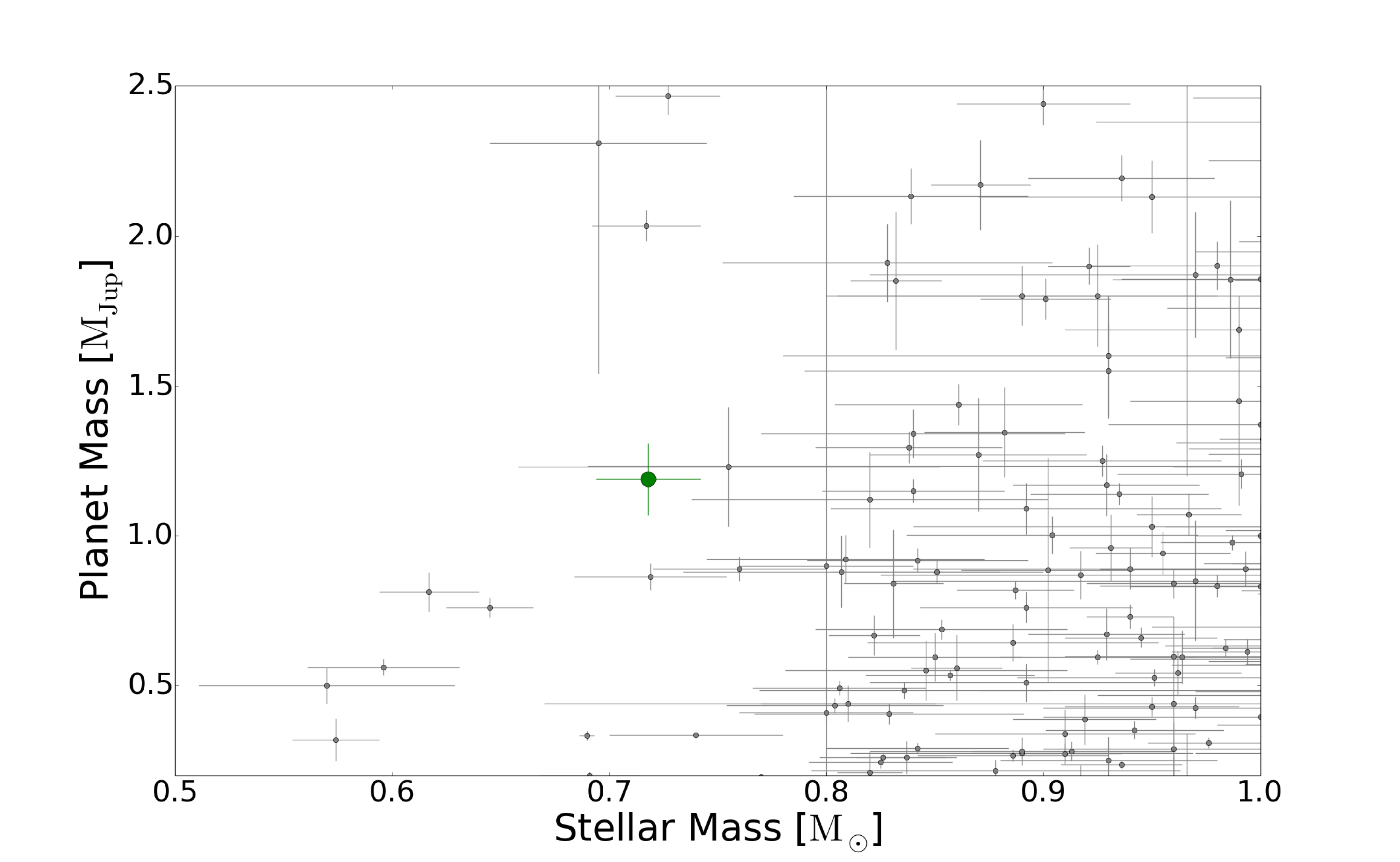}
\caption{Stellar mass versus planet mass, illustrating the scarcity of Jupiter- and 
super-Jupiter mass planets around low mass hosts $M_*<0.8\,\mathrm{M_{\odot}}$. Gray 
points are data from TEPCat, while the larger green point indicates Qatar-9b.}
\label{fig:mst_vs_mpl}
\end{figure}

An interesting aspect of Qatar-9b is the moderately large planet mass, $\mpl = 1.19 \mj$, in 
relation to the relatively low host star mass, $\mstar = 0.72 \msun$. Using data from 
TEPCat, we find 59 planets with host mass $\mstar < 0.8 \msun$ and only eight of these 
have masses in the range $0.7 \leq \mpl\,[\mj] \leq 2.5$. The position of Qatar-9b in this 
scarcely populated region of the parameter space is shown in Figure \ref{fig:mst_vs_mpl}.
We also note that Qatar-10b, at an orbital period of 1.65 days, is situated on the upper edge 
of the "sub-Jupiter desert'' \citep[e.g.][and see again 
Figure \ref{fig:porb_vs_mass}]{desert1,desert2}. 

The combination of relatively low surface gravity and relatively high equilibrium temperature 
for both Qatar-8b and Qatar-10b imply considerable-sized atmospheres. The atmospheric 
scale height, given by $H\,=\,kT/g \mu$ and assuming a Jupiter mean molecular mass of 2.3 
times the mass of a proton, is calculated to be $\sim$930\,km for Qatar-8b and $\sim$850\,km
for Qatar-10b. Taking into account the somewhat large host radii, the absorption signal, A, of 
an annular area of one atmospheric scale height during transit \citep{transmission1} is 
calculated to be $\sim$200\,ppm and $\sim$250\,ppm for Qatar-8b and 10b, respectively.



\section*{Acknowledgements}

This publication is supported by NPRP grant no. X-019-1-006 from the Qatar National 
Research Fund (a member of Qatar Foundation). The statements made herein are solely 
the responsibility of the authors. The Nanshan 1m telescope of XAO is supported by the CAS 
"Light of West China" program (XBBS-2014-25,2015-XBQN-A-02), and the Youth Innovation 
Promotion Association CAS (2014050). This article is partly based on observations made with 
the MuSCAT2 instrument, developed by ABC, at Telescopio Carlos S\'{a}nchez operated on 
the island of Tenerife by the IAC in the the Spanish Observatorio del Teide. This work is partly 
financed by the Spanish Ministry of Economics and Competitiveness through grants 
ESP2013-48391-C4-2-R. TRAPPIST-North is a project funded by the University of Li\`{e}ge, 
and performed in collaboration with Cadi Ayyad University of Marrakesh. The research leading 
to these results has received funding from an ARC grant for Concerted Research Actions 
financed by the Wallonia-Brussels Federation. M. Gillon and E. Jehin are F.R.S.-FNRS Senior 
Research Associates. L.M.\ acknowledges support from the Italian Minister of Instruction, 
University and Research (MIUR) through FFABR 2017 fund and from the University of Rome 
Tor Vergata through ``Mission: Sustainability 2016'' fund. This work has made use of data from 
the European Space Agency (ESA) mission {\it Gaia} 
(\href{https://www.cosmos.esa.int/gaia}{https://www.cosmos.esa.int/gaia}), processed by the 
{\it Gaia} Data Processing and Analysis Consortium (DPAC,
\href{https://www.cosmos.esa.int/web/gaia/dpac/consortium}{https://www.cosmos.esa.int/web/gaia/dpac/consortium}). 
Funding for the DPAC has been provided by national institutions, in particular the 
institutions participating in the {\it Gaia} Multilateral Agreement. We also acknowledge 
support from JSPS KAKENHI Grant Number no.\ JP16K13791, 17H04574, 18H01265 and 
18H05439, and JST PRESTO grant no.\ JPMJPR1775.




\end{document}